\documentclass[conference]{IEEEtran}
\IEEEoverridecommandlockouts

\usepackage[utf8]{inputenc} 
\usepackage[T1]{fontenc}    
\usepackage{hyperref}       
\usepackage{url}            
\usepackage{booktabs}       
\usepackage{amsfonts}       
\usepackage{nicefrac}       
\usepackage{microtype}      
\usepackage{lipsum}
\usepackage{graphicx, subcaption}
\usepackage{multirow, multicol, makecell}
\usepackage{xcolor}
\usepackage{algorithm2e}
\usepackage{subfiles}

\graphicspath{{images/}}

\def\BibTeX{{\rm B\kern-.05em{\sc i\kern-.025em b}\kern-.08em
    T\kern-.1667em\lower.7ex\hbox{E}\kern-.125emX}}

\begin{document}


\title{
Learning post-processing for QRS detection using Recurrent Neural Network
}

%



\author{\IEEEauthorblockN{Ahsan Habib}
\IEEEauthorblockA{\textit{School of Information Technology} \\
\textit{Deakin University}\\
Geelong, Australia 3225 \\
mahabib@deakin.edu.au \\
ORCID:0000-0002-1523-4251}
\and
\IEEEauthorblockN{Chandan Karmakar}
\IEEEauthorblockA{\textit{School of Information Technology} \\
\textit{Deakin University}\\
Geelong, Australia 3225 \\
karmakar@deakin.edu.au \\
ORCID:0000-0003-1814-0856}
\and
\IEEEauthorblockN{John Yearwood}
\IEEEauthorblockA{\textit{School of Information Technology} \\
\textit{Deakin University}\\
Geelong, Australia 3225 \\
john.yearwood@deakin.edu.au \\
ORDID:0000-0002-7562-6767}
}

\maketitle

\begin{abstract}

Deep-learning based QRS-detection algorithms often require an essential post-processing to refine the prediction-streams for R-peak localisation.
The post-processing performs signal-processing tasks from as simple as, removing isolated 0s or 1s in the prediction-stream to sophisticated steps, which require domain-specific knowledge, including the minimum threshold of a QRS-complex extent or R-R interval.
Often these thresholds vary among QRS-detection studies and are empirically determined for the target dataset, which may have implications if the target dataset differs.
Moreover, these studies, in general, fail to identify the relative strengths of deep-learning models and post-processing to weigh them appropriately.
This study classifies post-processing, as found in the QRS-detection literature, into two levels - moderate, and advanced - and advocates that the thresholds be learnt by an appropriate deep-learning module, called a Gated Recurrent Unit (GRU), to avoid explicitly setting post-processing thresholds.
This is done by utilising the same philosophy of shifting from hand-crafted feature-engineering to deep-learning-based feature-extraction.
The results suggest that GRU learns the post-processing level and the QRS detection performance using GRU-based post-processing marginally follows the domain-specific manual post-processing, without requiring usage of domain-specific threshold parameters.
To the best of our knowledge, the use of GRU to learn QRS-detection post-processing from CNN model generated prediction-streams is the first of its kind.
The outcome was used to recommend a modular design for a QRS-detection system, where the level of complexity of the CNN model and post-processing can be tuned based on the deployment environment.

\end{abstract}


\section{Introduction}

An electrocardiogram - also called ECG - records the electrical signals in one's heart.
It is a painless, and non-invasive way to help diagnose many common heart problems in people of all ages.
ECGs are often done in clinical settings, including a doctor's office, clinic or a hospital room, but now-a-days, smart wearable devices (i.e. smart watch) are commonly used for home-based cardiac monitoring.
ECG signals consist of waves - P-wave, QRS-complex, and T-wave, among which the QRS-complex is the most prominent and detection of which often forms the basis of automated ECG signal analysis.
The shift in ECG operating conditions from the clinical-setting to wearable devices, for cardiac-health monitoring, brings renewed focus to the QRS-detection research area which has been active for more than three decades.

Traditional QRS-detection algorithms generally employed two steps, firstly, a signal-processing step to enhance the QRS-complex, including the use of signal derivatives, digital filters, and signal transformation, and finally, the usage of amplitude-based thresholds to identify R-peaks \cite{Kohler2002}.
The QRS-detection literature contains classical machine-learning (ML) based approaches, which derive time or frequency-domain statistical signal properties to train models, including SVM \cite{mehta2008svm}, Random-Forest \cite{kropf2017ecg}, or KNN \cite{saini2013qrs} to identify whether an ECG segment contains a QRS-complex.
The traditional filter and classical ML based studies were often used with only a single or small number of datasets to report QRS-detection performance.
The study of Habib et al. \cite{habib_impact_2019} explored the characteristics of ECG-datasets and their uniqueness, in terms of noise and inter-patient variance, and found that a QRS-detection method performing well on a dataset, may not perform the same in other datasets.
Deep-learning based algorithms are encountered in the recent QRS-detection literature which show superior detection-scores for multiple validation datasets.

Deep-learning (DL) has become the de-facto standard for computer-vision tasks, which inspired QRS-detection researchers to follow the same path \cite{chandra_robust_2019, Xiang2018}.
Many of the DL-based QRS-detection studies were found to partition a single dataset to be used for training and validation respectively, until some of the recent studies, which were found to report superior performance across multiple validation datasets \cite{cai_qrs_2020, liu_octave_2020}.
These studies tend to create sophisticated DL model architecture, followed by an essential post-processing for finer R-peak localisation \cite{cai_qrs_2020, jia_high_2019, liu_octave_2020, sarlija_convolutional_2017, Lee2018, LEE201966}.
The post-processing was found to be used to remove isolated high or low predictions which may not form a legitimate QRS-candidate, as well as, to do advance signal-processing which may involve the use of domain-specific knowledge.
Based on the QRS-detection and R-peak localisation literature, the post-processing that requires domain-specific knowledge was divided into two levels, firstly, a moderate level which performs signal-processing based on some domain-specific heuristic, including removal of a candidate QRS-complex \cite{cai_qrs_2020, Lee2018, LEE201966} or QRS-complex cluster \cite{sarlija_convolutional_2017} which falls below an impossiblly small QRS-complex duration threshold, and secondly, an advanced level which may involve further signal-processing assuming a minimum distance between R-peaks for two consecutive candidate QRS-complexes \cite{jia_high_2019, LEE201966}.
An essential salt-and-pepper filter, which merely removes isolated sample predictions (i.e. remove isolated 0s or 1s in a binary prediction-stream \cite{jia_high_2019}), is often required for the mentioned moderate, and advanced post-processing.
QRS-detection studies claim superior detection performance using sophisticated and deep models and moderate/advanced post-processing, but generally not found to weigh them separately to clearly understand which one contributes and to what extent.

Post-processing in QRS-detection studies often assume different thresholds, for example, the minimum R-R distance of 200 milli-seconds worked well in our study, while another study used 100 milli-seconds \cite{cai_qrs_2020}.
This threshold is likely not to be set following a rule-of-thumb, rather would be determined empirically based on the target datasets.
It can be argued that since the thresholds were determined empirically from the dataset, it would be preferable to let a model learn it from the very dataset itself.
A post-processing model should learn parameters like the minimum extent of candidate QRS-complex or R-R distance from the data.
These parameters relate the temporal relationship of QRS-complexes in ECG data and recurrent-neural-networks (RNN) are primarily used to unveil temporal sequences.
Following the inherent nature of RNN, a light-weight counterpart, called Gated Recurrent Unit (GRU), was used to learn the post-processing.

This study aims to train a GRU to learn data-driven post-processing, based on solely the CNN model generated binary prediction-streams, to remove the subjectivity (use of domain knowledge, dataset specific design etc.) from post-processing reported in literature.
The primary focus of this study was not to find the best CNN model for QRS detection.
Since shallow CNN models are often utilised in QRS detection~\cite{chandra_robust_2019,Xiang2018,sarlija_convolutional_2017}, unlike the trend of using deep models for computer-vision tasks,
a shallow 2-layer convolution-only CNN model was defined as the baseline-convnet and its complexity was increased by increasing the depth to 4, 8, 16, 32, and 64-layer.
For the post-processing, we used a single GRU module and optimised its two hyperparameters - the number of hidden-layers and sequence-length.
The CNN models (baseline-convnet and deeper variants) were trained to generate binary prediction-streams and the GRU was trained to post-process the prediction-streams for further refinement.
The performance of the CNN-GRU model was compared with the CNN-traditional post-processing (two levels namely moderate, and advanced) for seven cross-validated datasets.
The outcome was used to recommend a modular design concept for a QRS-detection system, where the level of complexity of the CNN model and post-processing can be tuned based on the deployment environment.

\section{Related Works}

The use of deep-learning-based models are common in contemporary QRS-detection studies in the literature.
A multi-branch dilated six-layer deep CNN model, with a SENet before the final decision layer, used a post-processing step to revise the prediction-stream through multiple iterations~\cite{cai_qrs_2020}.
In a separate study, a similar multi-branch dilated convolution was used, followed by a post-processing to remove isolated 0s and 1s from the output prediction-stream of the CNN-model~\cite{jia_high_2019}.
There are studies that use - different types of convolution~\cite{liu_octave_2020}, multi-channel ECG~\cite{Lee2018,LEE201966}, and fully-convolutional networks~\cite{LEE201966}, where all of them consist of an essential manual post-processing step, which utilises domain-specific knowledge, at some level, to refine the deep-learning-based model-generated prediction-stream to finally localise R-peaks.
The idea of automating the post-processing task using a deep-learning-based module has not been explored yet, rather the focus has been on the use of a sophisticated deep-learning-based model with a manual post-processing.

\section{Methodology} \label{method}

\subsection{Problem Formulation}
There are two tasks in this study and each uses the corresponding model to solve its specific problem.
The first task is QRS-prediction and the other is learning post-processing.
The QRS-prediction task was formulated as a segmentation-problem where each input sample receives an output binary-prediction from a CNN model, whether it belongs to a QRS-region, in case the prediction score is above a threshold, or a non-QRS region.
During model training, each annotated R-peak was labeled as 1s around 0.05 seconds of equivalent sample on each side, forming a region of five samples (considering 100Hz ECG signal) as a representative QRS-region, while all other samples (as 0s), form the non-QRS regions.

The post-processing learning was formulated as a sequence-learning task using a Gated Recurrent Unit (GRU), a variant of an RNN, which was trained to generate a binary-sequence similar to the annotated QRS/non-QRS binary-sequence from the binary prediction-stream of the CNN model.
This task essentially makes a GRU learn to repair a prediction-stream making the post-processing redundant, which requires domain-specific-heuristic based signal-processing (i.e. moderate, and advanced post-process).
Note that the post-processing levels are described in Methodology section.
The smoothed out prediction-stream is subject to salt-and-pepper filtration followed by R-peak localisation.

\subsection{ECG Data}
Eight ECG datasets from the PhysioNet \cite{goldberger_physiobank_2000} databank were used in this study, including the MIT-BIH-Arrhythmia \cite{moody_impact_2001}, INCART, QT \cite{laguna_database_1997}, EDB (European ST-T Database) \cite{taddei_european_1992}, STDB (MIT-BIH ST Change Database), TWADB (T-Wave Alternans Challenge Database), NSTDB (MIT-BIH Noise Stress Test Database) \cite{moody_noise_1984}, and SVDB (MIT-BIH Supraventricular Arrhythmia Database) \cite{greenwald_improved_1990}.
Each of these databases are summarised in Table \ref{tab:tbl_databases}.
The primary (or first) lead from each of the datasets were used in this study, and valid beat types include, \emph{NLRBAJSVFREQ/}, which avoided paced-beat.

\begin{table}[h]
    \centering
    \caption{Characteristics of PhysioNet datasets used in current study.}
    \begin{tabular}{|c|c|c|c|c|c|}
      \hline
      \emph{DB-Name} & \emph{\makecell{Source \\Hz}} & \emph{\#Rec.} & \emph{\makecell{Used \\\#Rec.}} & \emph{\makecell{Len \\(minute)}} & \emph{\#Beats} \\
      \hline
      MIT-BIH-Arr & 360 & 48 & 45 & 30 & 102941 \\
      \hline
      INCART & 257 & 75 & 75 & 30 & 175900 \\
      \hline
      QT & 250 & 82 & 80 & 15 & 84883 \\
      \hline
      EDB & 250 & 90 & 90 & 120 & 791665 \\
      \hline
      STDB & 360 & 28 & 28 & vary & 76175 \\
      \hline
      TWADB & 500 & 100 & 100 & 2 & 18993 \\
      \hline
      NSTDB & 360 & 12 & 12 & 30 & 25590 \\
      \hline
      SVDB & 128 & 78 & 78 & 30 & 184583 \\
      \hline
    \end{tabular}
    \label{tab:tbl_databases}
\end{table}

\subsection{Data Pre-processing}
\paragraph{Signal Resampling}
In this study, we used eight datasets from Physionet, which are unique in terms of noise, sampling frequency, and inter-patient variance \cite{habib_impact_2019}.
For usability, each dataset was re-sampled at a single common frequency of as low as 100Hz, considering a domain-specific heuristic that an average QRS-complex may go upto 25Hz or beyond \cite{ajdaraga_analysis_2017}.
Thus, the Nyquest sampling frequency \cite{proakis_digital_2004} should be at least 50Hz to contain an average QRS-complex.
Considering the effect of noise and QRS-complex extremes, a sample frequency of 100Hz was adopted, which also benefits maintaining a small deep-learning model size.
Since all datasets were from PhysioNet databank, a suitable utility program \emph{xform} was used to resample both the signal and annotation.

\paragraph{Segmentation and Overlapping}
The QRS-segment-classification problem restricts a segment length to be able to contain at most a single QRS-complex per segment.
Our study was formulated as a sample-wise binary-segmentation problem, where each input sample receives an output prediction, and thus, any number of QRS-complexes may be used to form segments.
To keep the deep-learning model small, ECG records were sliced into three second segments with two seconds overlapping.
A two seconds overlapping strategy was likely to increase the detection probability, since each QRS-complex goes through the model three times, which were then aggregated in the binary decision-level using a logical OR operation.
The segmentation and overlapping strategies were used following a similar strategy explained in another study \cite{jia_high_2019}.

\paragraph{Normalisation}
We have applied segment-wise Z-score normalisation to subtract the population mean.
Other normalisation, like Min-Max was found inappropriate for our study, since it cannot handle the variation due to mean and outliers.

\subsection{Deep-learning Models}
This study examines the possibility of the usage of recurrent neural-networks in learning the post-processing so that advanced signal-processing and domain-specific knowledge can be avoided as much as possible.
A CNN model was used to input z-score normalised single-channel ECG signals and to generate sample-wise prediction indicating whether a sample belongs to a QRS-region or not.
Since finding the best model is not the focus of this study, a simple and shallow 2-layer convolution-only CNN model was used to generate the binary prediction-stream, which would be referred to as the baseline-convnet or baseline-network in the text.
Unlike computer-vision tasks, where using a deep network is common, the use of a single-\cite{chandra_robust_2019} or two-convolution-layer~\cite{Xiang2018,sarlija_convolutional_2017} CNN models for detecting beat in ECG time-series signal is often found beneficial.
The deep-learning-based models, i.e., convolutional neural network and GRU, were implemented in Python 3.7 using PyTorch 1.8.1 API.

\begin{figure*}[h]
  \centering
  \begin{subfigure}[b]{0.61\textwidth}
    \includegraphics[width=\textwidth]{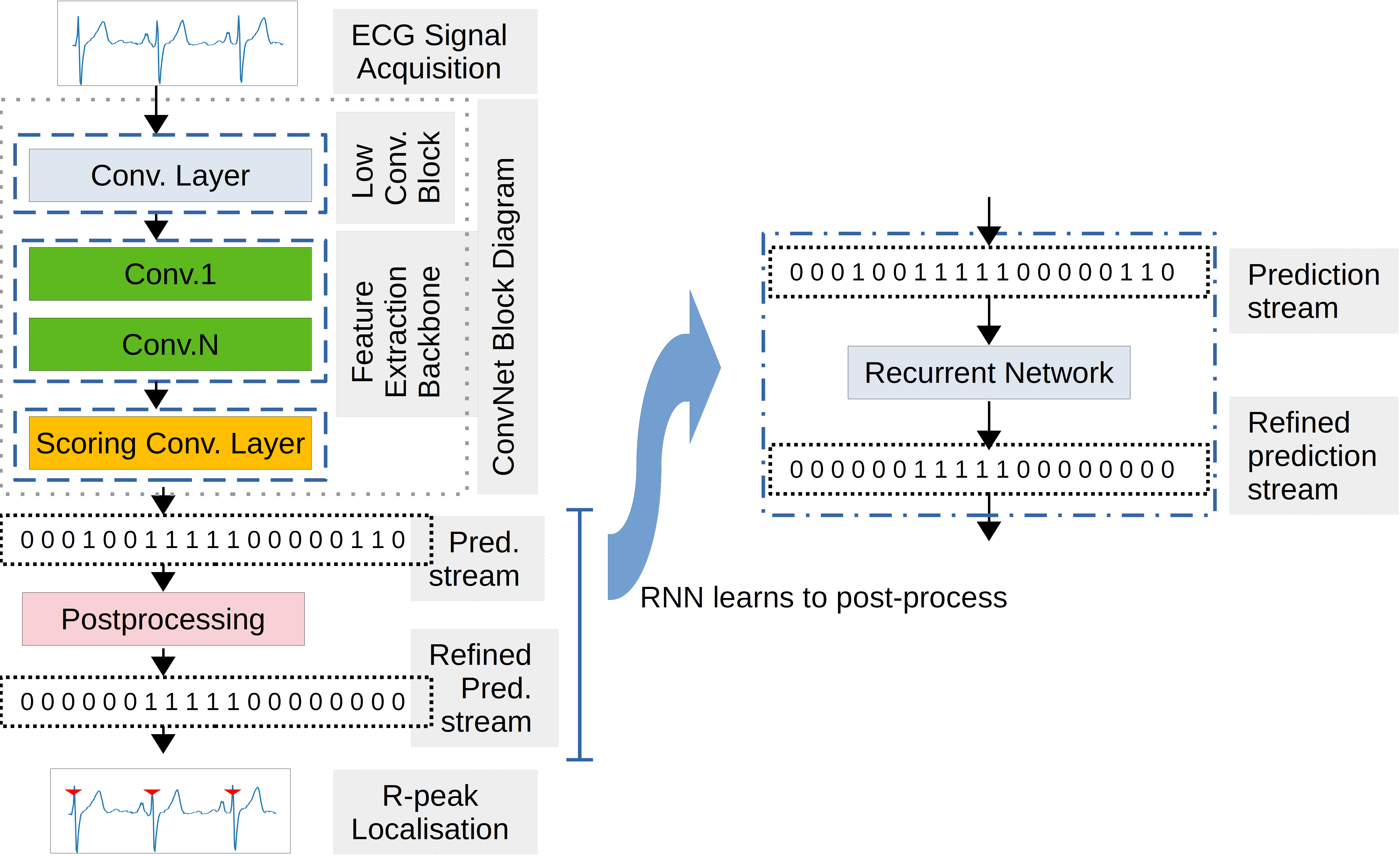}
  \end{subfigure}

  \caption{(Left) Convolution-only network block-diagram, which was used to generate sample-wise prediction, from single-channel raw ECG signal, a high-probability if a sample belongs to a QRS-complex, low-probability otherwise. (Right) A recurrent neural network (RNN) learns to repair the binary prediction-stream, generated by the CNN model, which then followed by salt-and-pepper filtration and R-peak localisation. RNN was expected to learn the temporal relationship in the prediction-stream to make the moderate and advanced levels of post-processing redundant.}
  \label{fig:network}
\end{figure*}

Figure \ref{fig:network} shows, on the left, the convolutional-network block diagram highlighting the low-convolution-block, feature-extracting-backbone, and scoring layer.
The complexity of the baseline-convnet was increased by adding more convolution-layers, including 4, 8, 16, 32, and 64-layer to observe the performance improvement.
Convolution kernel-size is one of the hyper-parameters, to be optimised, however, a domain-specific heuristic was used to achieve this, following the study in \cite{guo_inter-patient_2019}. 
This study found that since the average QRS-complex is 60 milli-seconds, a kernel-size of 44 milli-seconds should capture most of its components.
For a 100Hz ECG signal, the 44 milli-seconds yields a convolution-kernel of 5 samples long.
Being a sample-wise segmentation problem, there are CNN models in the segmentation literature, including U-Net \cite{navab_u-net_2015} that uses sub-sampling layers at earlier stages, followed by an up-sampling operation at later-stages to re-construct the output resolution to be the same as the input.
In this study, the use of sub-sampling layers were avoided completely to keep the model simple and avoid any re-construction effort, - following a network-design philosophy explained in the study in~\cite{pelt_mixed-scale_2018}.
The number of output channels was set to a fixed number 24 for all the convolution layers, since very few channels were found to degrade the detection performance, while more channels were found unable to gain performance proportionately but increased the complexity.
The output feature-map were batch-normalised (BN) before passing to the ReLU non-linear activation function, following the argument that BN accelerates training, enables higher learning-rate, and improves generalisation accuracy \cite{ioffe2015batch}.
The CNN model generates a binary prediction-stream which goes through essential multi-level post-processing involving signal-processing and domain-specific heuristics.

To facilitate deep-learning-based post-processing, the Gated Recurrent Unit (GRU) \cite{cho2014learning}, which is a light-weight (i.e. reduced number of gates) and faster sibling of Long Short-Term Memory (LSTM), was used, as shown in the right-portion of Figure \ref{fig:network}.
GRUs can be stacked to optimise for a particular task, however, here a GRU with a single and double hidden-layer were explored, with a range of sequence-lengths (i.e. 1-5 seconds), to find an optimum GRU configuration.

\subsection{Post-processing Levels}
Post-processing is often essential for R-peak localisation and the level of post-processing adopted in studies varies.
Based on the signal-processing logic and domain-specific knowledge utilised, the post-processing in the QRS-detection literature was categorised into two levels - moderate, and advanced.
A primitive signal-processing is often essential for other post-processing to remove isolated 1s or 0s.
In this study, the salt-and-pepper noise filter has been considered as a primitive step that aggregates consecutive 1s in CNN model generated binary prediction-streams and remove isolated 1s or 0s.

\begin{algorithm}
\caption{salt-and-pepper filter} \label{algo:postprocessing_minimal}
\DontPrintSemicolon
\KwData{CNN model's binary prediction-stream (100Hz signal)}
\KwResult{Refined prediction-stream}
\SetKwInput{Kw}{Salt-and-pepper filter} \Kw{}
\SetKwBlock{Begin}{begin}{end} \Begin{
  \nl \emph{Group 1s:} Scan pred-stream to locate consecutive 1s, create a Node structure
  with attributes - $start\_loc$, $confidence$ (i.e. number of 1s in a node), and $q\_loc$ (=start\_loc + 0.5*$confidence$) to prepare a node\_list.\;
  \nl \emph{Remove salt-and-pepper noise:} Scan $node\_list$ and merge two consecutive nodes not more than 3 samples apart.\;
}
\end{algorithm}

The salt-and-pepper filter, shown in Algorithm \ref{algo:postprocessing_minimal}, is thus defined to be a basic signal-processing step which groups the consecutive 1s to represent a QRS-complex node with certain information, including start-location, the number of consecutive 1s (a.k.a. confidence-score of a node), and candidate R-peak location (equals to start-location plus half of the confidence-score).
A list of nodes were formed by identifying and chaining the nodes for each ECG record.
Removal of salt-and-pepper noise, semantically, was carried out by identifying two close nodes which are three or fewer samples apart, which were then merged to form a more confident node.

\begin{algorithm}
\caption{Moderate postprocessing} \label{algo:postprocessing_moderate}
\DontPrintSemicolon
\KwData{CNN model's binary prediction-stream (100Hz signal)}
\KwResult{Refined prediction-stream}
\SetKwInput{Kw}{Moderate Postprocess} \Kw{}
\SetKwBlock{Begin}{begin}{end} \Begin{
  \nl \emph{Do salt-and-pepper filtering}\;
  \nl \emph{Filter less-confident nodes:} Scan $node\_list$ and remove nodes with $confidence$ value less than six (i.e. 64 milli-seconds equivalent threshold).\;
}
\end{algorithm}

The moderate post-processing, shown in Algorithm \ref{algo:postprocessing_moderate}, initially uses the essential salt-and-pepper filtering to form a list of candidate QRS-complex nodes and remove salt-and-pepper noise, followed by a step to filter less-confident nodes from the candidate node-list.
The filter operation is based on the logic that each node's confidence value should be at least 64 milli-seconds of equivalent number of samples, which is six samples for 100Hz signal.
These less-confident nodes were assumed to be the reason for a QRS-like artifact, however, this step may also remove some of the legitimate candidates, which may be subject to further steps of repair, but overlooked in our study.

\begin{algorithm}
\caption{Advanced postprocessing} \label{algo:postprocessing_advanced}
\DontPrintSemicolon
\KwData{(i) CNN model's binary prediction-stream (100Hz signal), (ii) first derivative of original ECG signal}
\KwResult{Refined prediction-stream}
\SetKwInput{Kw}{Advanced Postprocess} \Kw{}
\SetKwBlock{Begin}{begin}{end} \Begin{
  \nl \emph{Do moderate postprocess}\;
  \nl \emph{Filter low R-R interval nodes:} Scan final confident $node\_list$ and remove the right node of a current node if it is less than 200 milli-seconds apart.\;
  \nl During above R-R interval scan, make sure each node contains at least a sample with amplitude higher than quarter of the max amplitude within the corresponding first-derivative segment.\;
}
\end{algorithm}

The advanced post-processing, shown in Algorithm \ref{algo:postprocessing_advanced}, initially uses the moderate post-processing (which in-turn employs the essential salt-and-pepper filtering) to produce a confident list of QRS-complex candidates, followed by a further candidate-filtration step which calculates the distance between the candidate R-peaks of two consecutive nodes.
Recall that the candidate R-peak of each node was calculated by adding half of the extent of the consecutive 1s (i.e. half of the confidence score) with the start-location of candidate QRS-complex.
A very low R-R distance threshold of 200 milli-seconds (which is 20 samples for 100Hz signal) was used to remove candidate QRS-nodes.
The choice of a 200 milli-second threshold may seem to be high enough compared to the threshold set to 100 milli-seconds in another study \cite{cai_qrs_2020}, but the larger threshold set to our study worked as expected.
An additional attribute per QRS-complex candidate node was calculated, called \emph{support score}, which is the number of 1s in the first-derivative of the corresponding source signal-fragment with magnitude prominently high.
Each confident QRS-node that falls below the minimum R-R distance threshold is retained only if its support score is one or above, otherwise, the candidate node is dropped.
Further post-processing steps were identified in QRS-detection studies, for example, to recover less-confident candidate nodes that were removed during the moderate post-processing, by identifying if the R-R distance is beyond an impractically large threshold and allowing nodes with a slightly lower confidence-score \cite{cai_qrs_2020}.
However, our study limits the advanced post-processing to the usage of the above-mentioned mininum R-R threshold-based filtration to focus on the current study goal, which is to verify if RNNs can learn the post-processing to make the moderate and advanced post-processing, as defined in Algorithm \ref{algo:postprocessing_moderate}, and \ref{algo:postprocessing_advanced}, redundant.

\subsection{Model Training and Testing}

\subsubsection{CNN Model Training}

\begin{figure}[h]
  \centering
  \begin{subfigure}[b]{0.49\textwidth}
    \includegraphics[width=\textwidth]{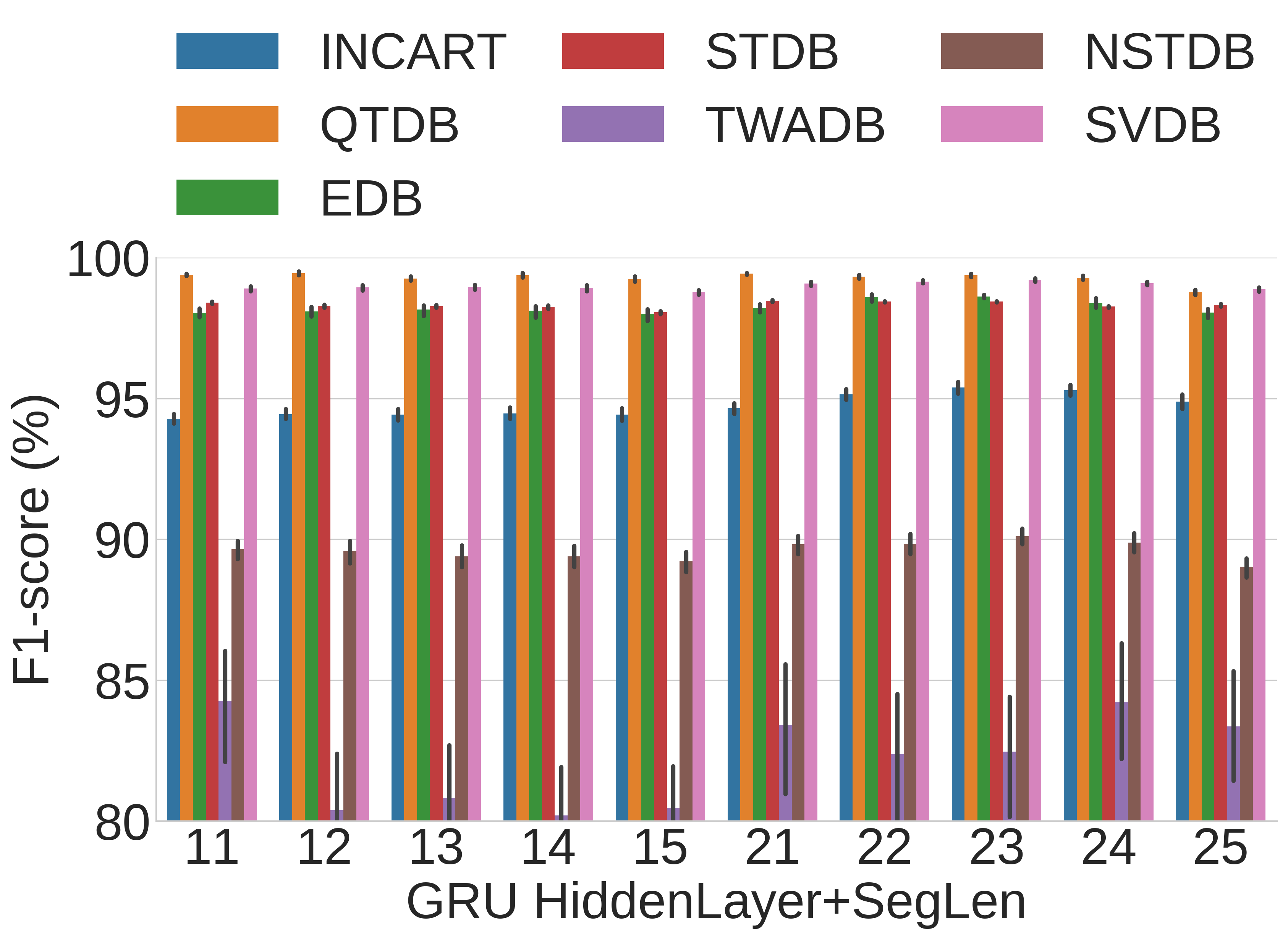}
    \caption{}
  \end{subfigure}
  \begin{subfigure}[b]{0.49\textwidth}
    \includegraphics[width=\textwidth]{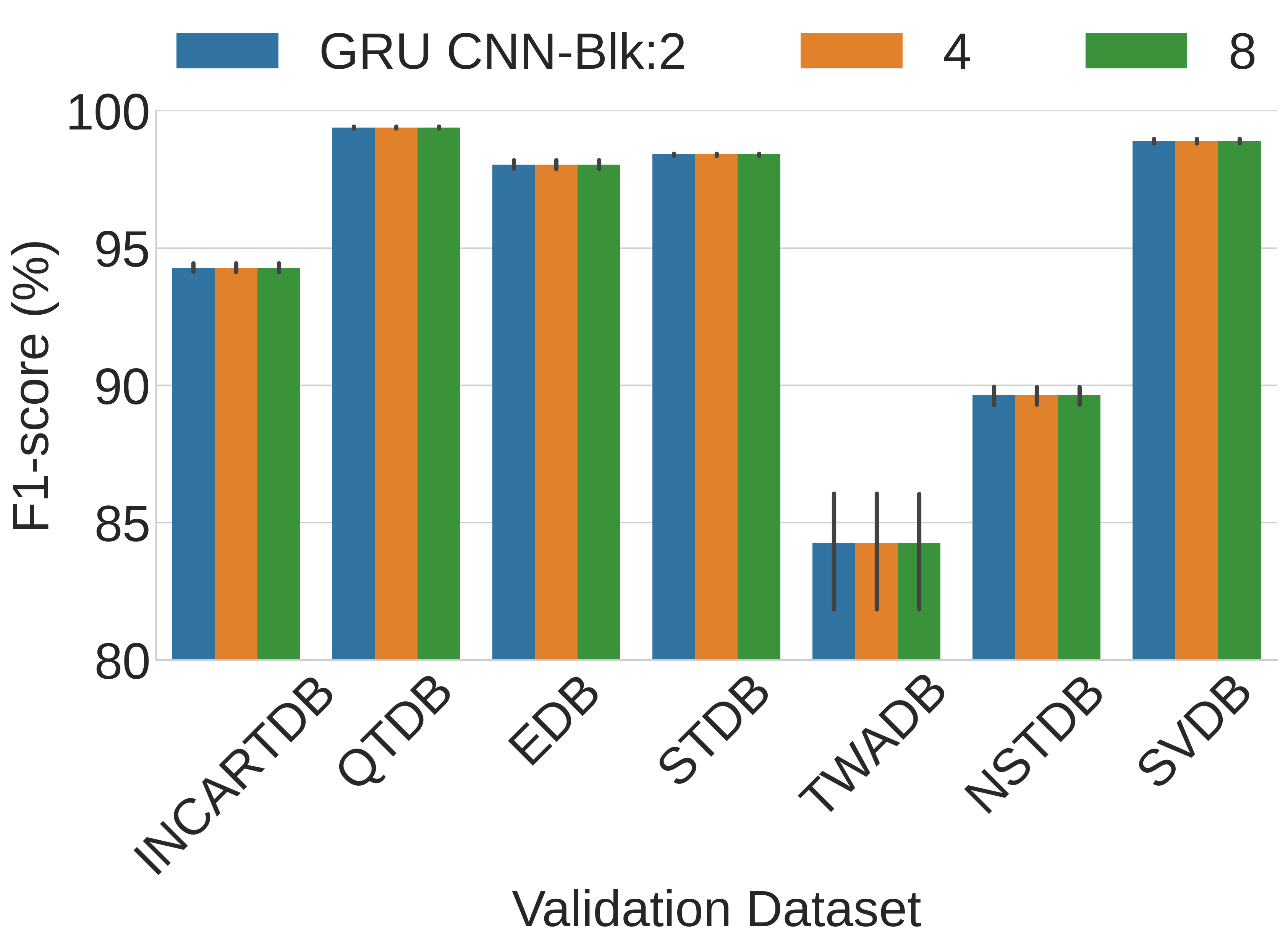}
    \caption{}
  \end{subfigure}

  \caption{GRU parameter tuning process. (a) F1-score variation for GRU's number of hidden-layer and sequence-length variation across validation-datasets using baseline-convnet to generate prediction-streams. The number of hidden-layer and sequence-length were encoded using a single two-digit number, where the first-digit indicates the number of hidden-layer and the second-digit the sequence-length in seconds. (b) F1-score variation of validation-datasets using GRU variants of a single hidden-layer and 1-second sequence-length, trained with binary prediction-stream generated from CNN models of 2, 4, and 8-layer deep network. It was hypothesized that GRUs, trained with shallow-network's noisy prediction-stream, may learn differently, but the performance was found to vary marginally.}
  \label{fig:gru_hyperparam}
\end{figure}

The baseline-convnet, as well as its deeper variants, were trained with the comparatively healthy-and-clean MIT-BIH-Arrhythmia dataset to facilitate learning QRS-complex morphology.
Subject-wise five-fold internal-validation was used, where in each iteration, four-fold worth recordings were segmented and used for model training, while the segments from the rest-fold were used to validate the model (by calculating validation loss).
A maximum of 100 epochs were set for the training with an early stopping mechanism was in place to verify if the validation-loss stalls for N=7 (selected empirically) consecutive epochs.
An eager learning was initially intended by setting a high learning-rate of 0.01, but a scheduler (i.e. LR-scheduler) was employed to regulate the learning-rate once validation-loss stops decreasing for five consecutive epochs.
It was found beneficial (from observations) to reduce the learning-rate by a factor of 0.1 once the development of validation-loss ceases for five consecutive epochs.
The slightly higher number of epochs set for the early-stopping module was intended to allow enough time for the LR-scheduler to decrease the learning-rate, so that during the next two epochs validation-loss becomes noticeable, if the model-training has not already reached the global-minima (or one of the local minimums).

\subsubsection{Testing Approach}
The trained CNN models were used to perform cross-database validation with two post-processing scenarios, (i) traditional post-processing levels (i.e. moderate, and advanced as described in Algorithm \ref{algo:postprocessing_moderate}, and \ref{algo:postprocessing_advanced}), as shown in Figure \ref{fig:network}-left-column, and (ii) GRU post-processing (Figure \ref{fig:network}-right-column)
The validation steps, irrespective of means of post-processing, are summarised as below -
\begin{itemize}
  \item With the five-fold training, five CNN models were produced with the CNN model training method. 
  The validation score for each test dataset was calculated using all five models.
  Validation scores of five models were aggregated and mean and standard deviation used for reporting in the results.
  \item This process was repeated for all variants (depth ranging $2^{1...6}$) of the baseline-convnet models.
\end{itemize}


\subsubsection{Training GRU for Post-processing}
%

To ensure that the GRU training data i.e., output binary stream from CNN model, includes an adequate level of noise, the following two conditions need to be satisfied: (i) the selected dataset should be inherently noisy; and (ii) the CNN model should be shallow, so that it makes lot of erroneous predictions.
In this study, the choice of the baseline-convnet (depth=2) was adopted as a shallow CNN model after empirical evaluation of layer 4 and 8 (see Figure \ref{fig:gru_hyperparam}-b).
INCART was selected as a dataset, which contains comparatively noisy ECG subjects consistent with ischemia, coronary artery disease, conduction abnormalities, and arrhythmias \cite{goldberger_physiobank_2000}.
Only the baseline-convnet generated prediction-stream of the INCART (instead of the ECG signal) was used for GRU training and validation.
In the GRU training, the number of hidden-layers were varied in the range of 1 to 2, while the sequence-length was varied in the range of 1 to 5 seconds.
A similar training strategy of the baseline-convnet (and deep variants) was used for the GRU training, where an early-stopping was used with a patience score of seven and an LR-scheduler with a patience score of five.
Usage of an initial learning-rate of 0.01, that was used in the CNN model training, was found inconsistent and a smaller initial LR of 0.001 was used.
The GRU was found to require more epochs to converge, may be due to the usage of a smaller initial learning-rate, the maximum epoch of 200 was set for the GRU training.

\subsubsection{GRU Post-process Validation}
The output binary prediction-stream of all CNN model variants were passed through the trained GRU model for post-processing.
A salt-and-pepper filter was used on the output of GRU model followed by R-peak localisation and calculation of validation score.

\subsection {Validation score}
In this study, an F1-score was used as a validation score for proposed models.
F1-score is calculated as
\begin{equation}
F1 = 2 * \frac {PPV * Sensitivy} {PPV + Sensitivy}
\end{equation}


\section{Results} \label{results}

\begin{figure*}[h!]
  \centering
  \begin{subfigure}[b]{0.32\textwidth}
    \includegraphics[width=\textwidth]{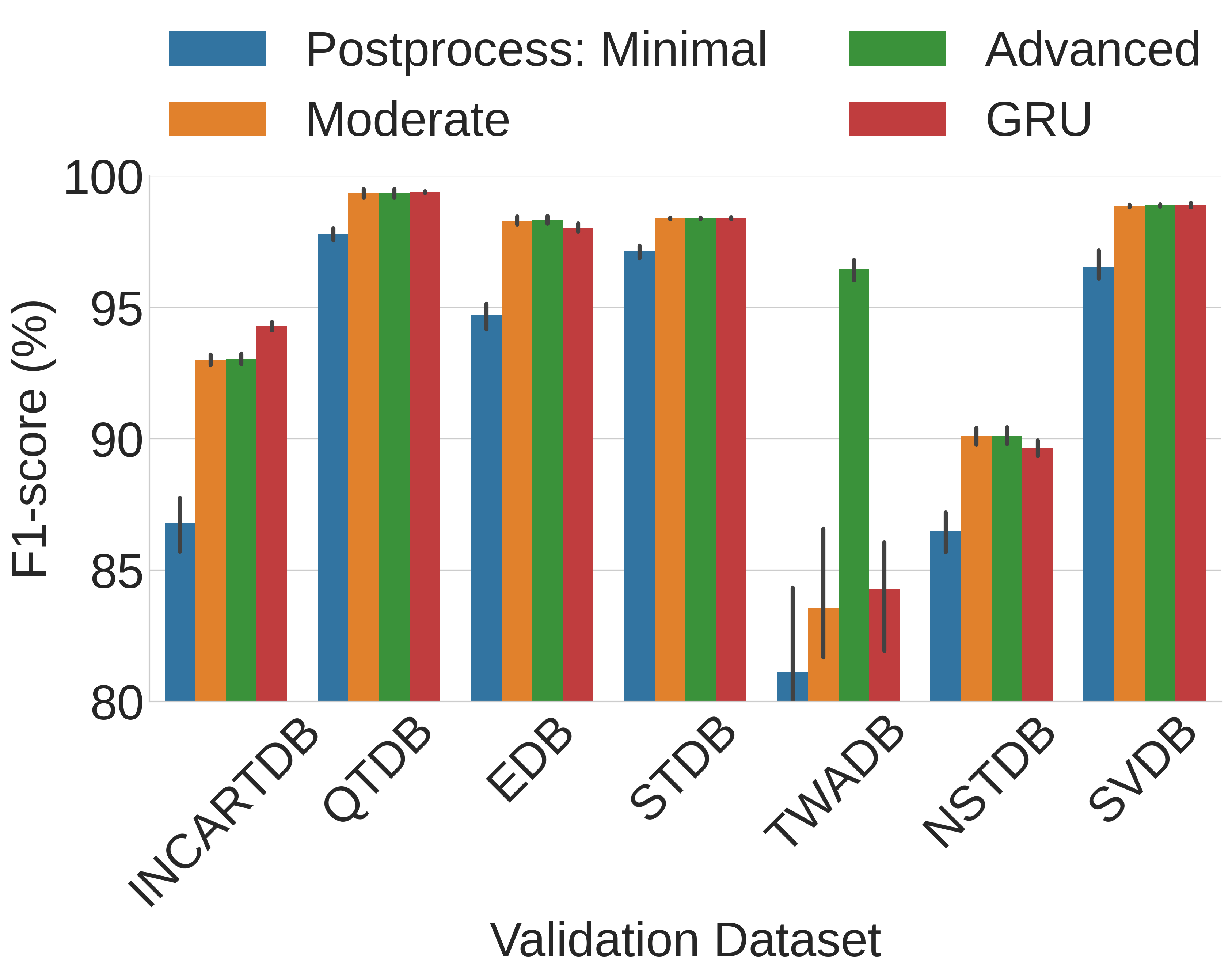}
    \caption{}
  \end{subfigure}
  \begin{subfigure}[b]{0.32\textwidth}
    \includegraphics[width=\textwidth]{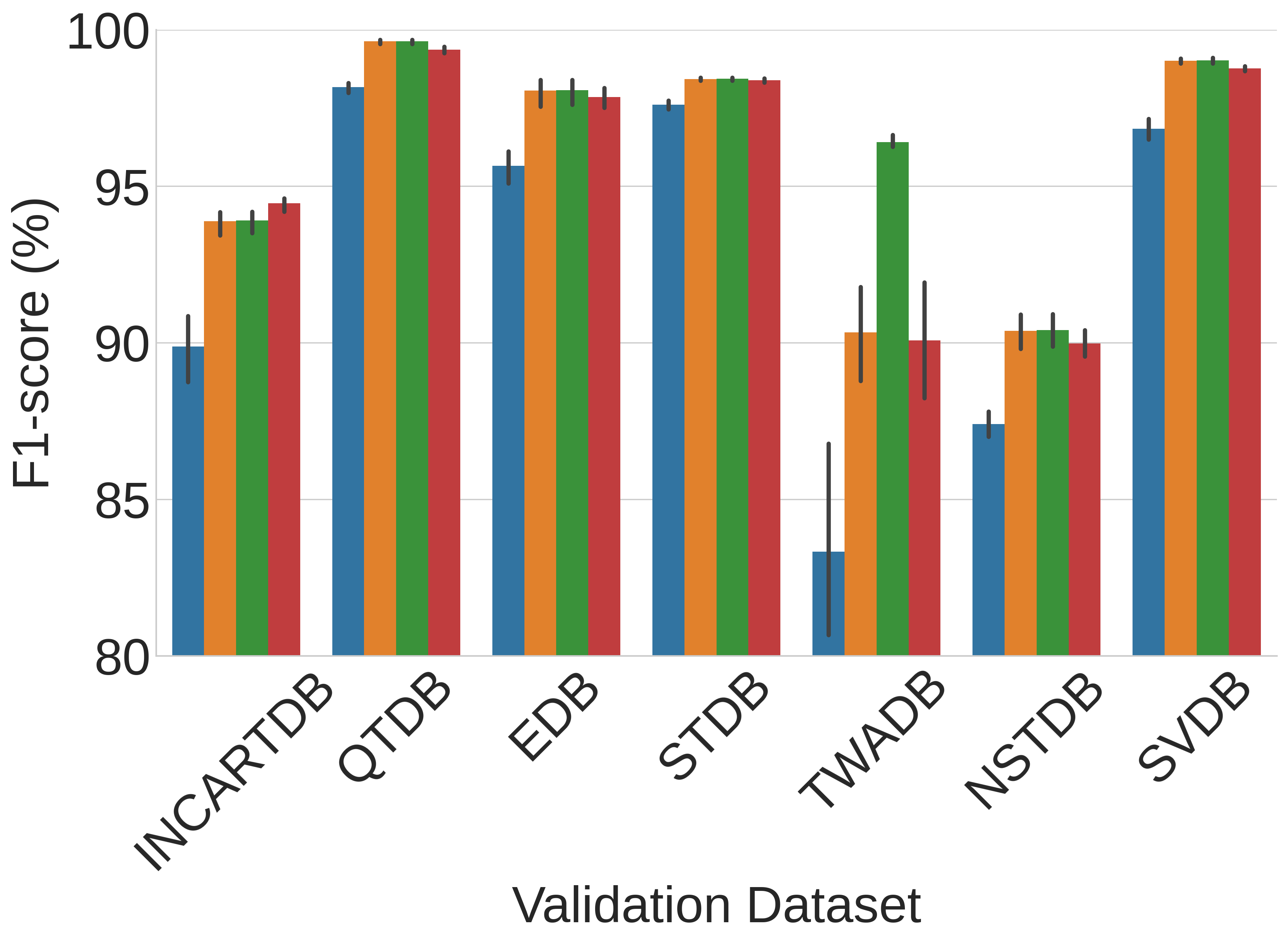}
    \caption{}
  \end{subfigure}
  \begin{subfigure}[b]{0.32\textwidth}
    \includegraphics[width=\textwidth]{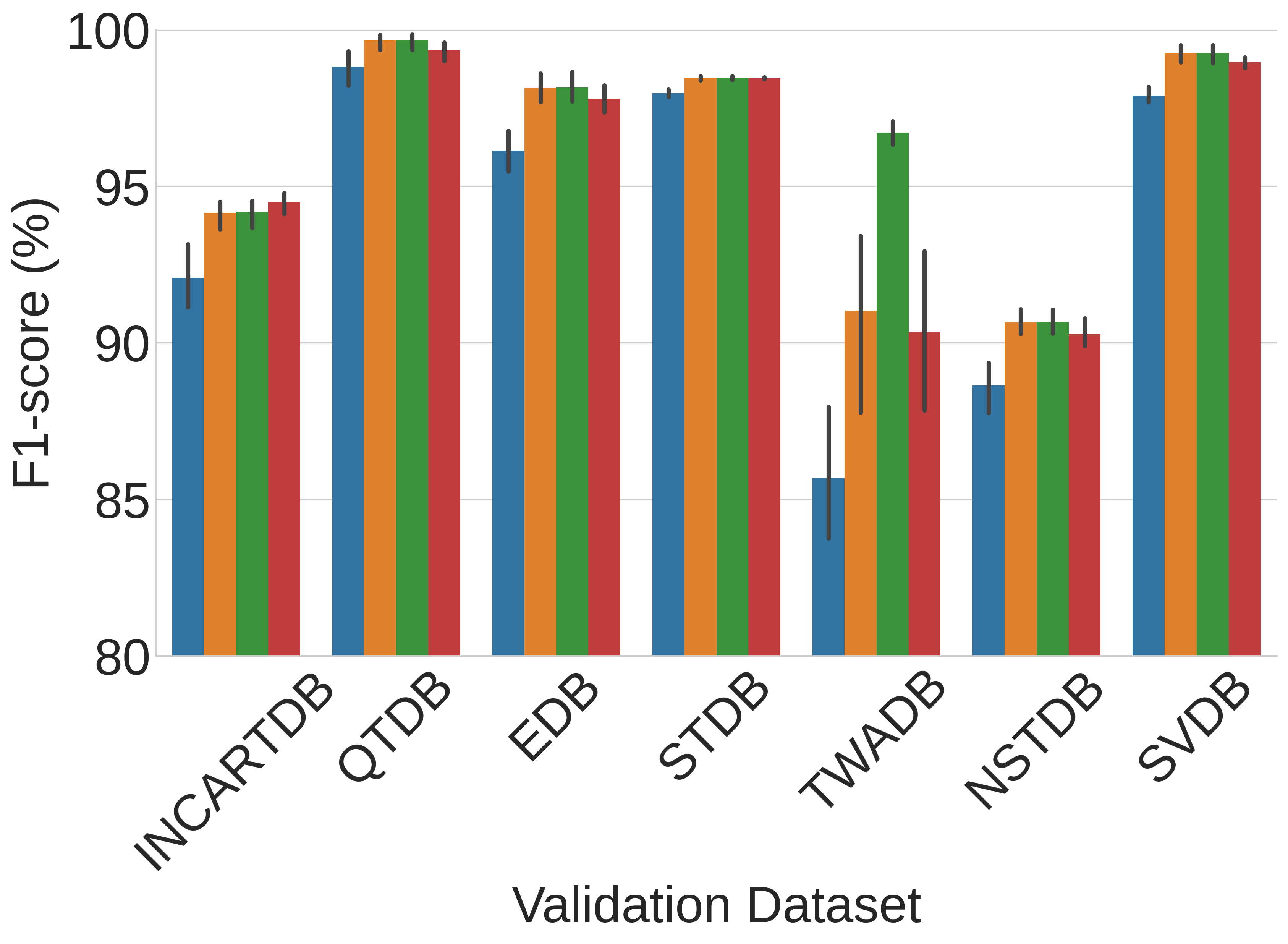}
    \caption{}
  \end{subfigure}
  \begin{subfigure}[b]{0.32\textwidth}
    \includegraphics[width=\textwidth]{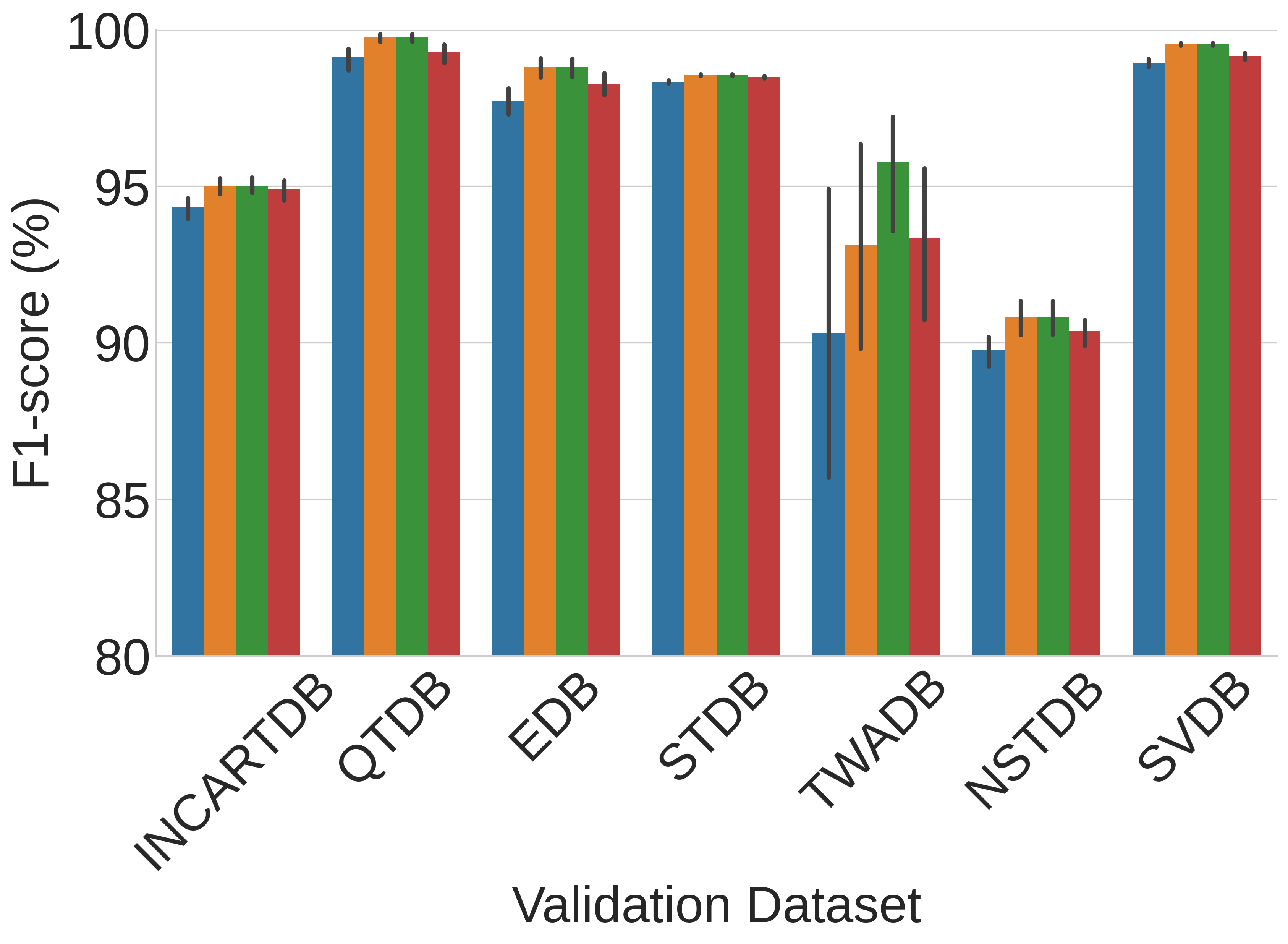}
    \caption{}
  \end{subfigure}
  \begin{subfigure}[b]{0.32\textwidth}
    \includegraphics[width=\textwidth]{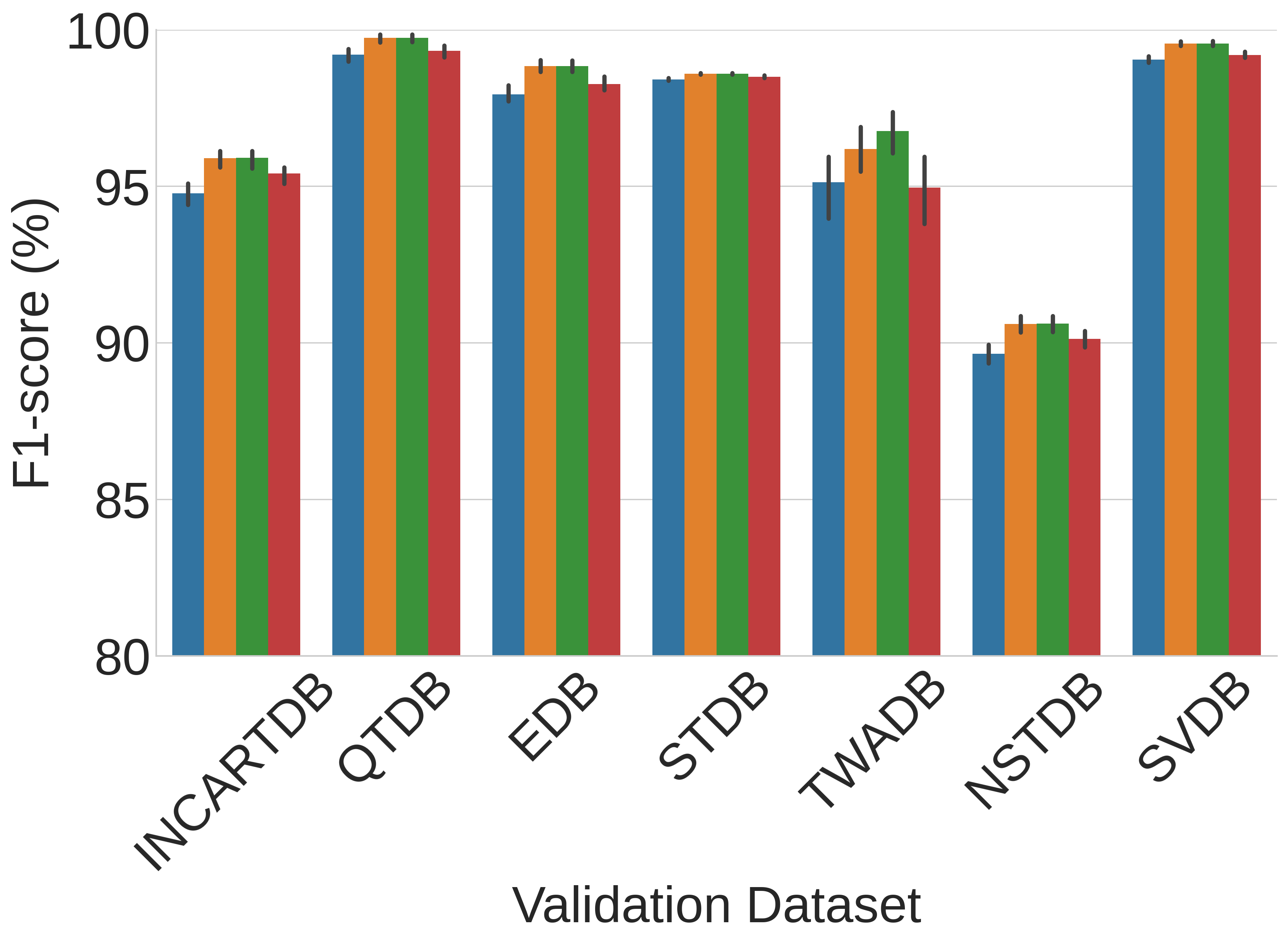}
    \caption{}
  \end{subfigure}
  \begin{subfigure}[b]{0.32\textwidth}
    \includegraphics[width=\textwidth]{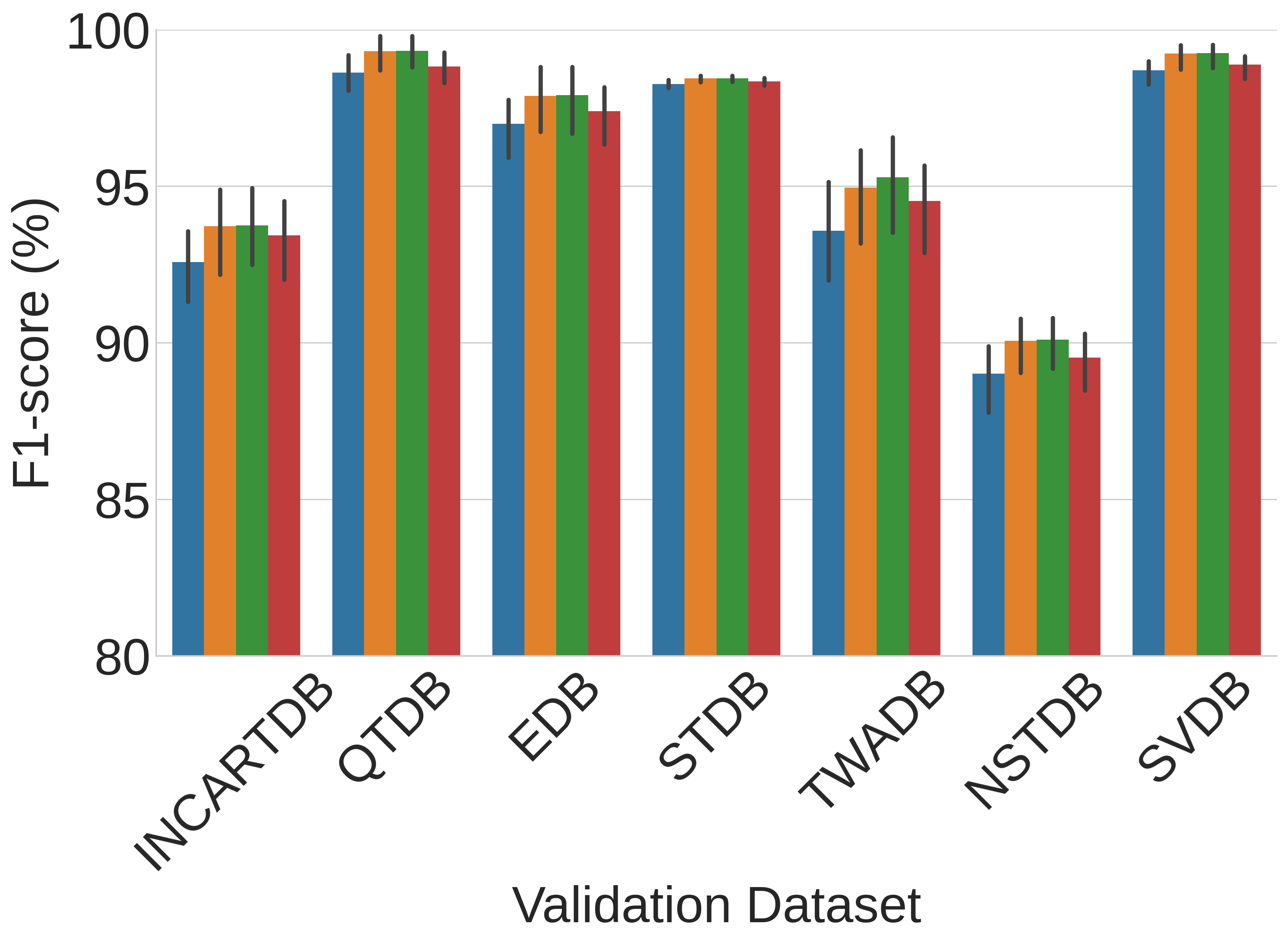}
    \caption{}
  \end{subfigure}

  \caption{F1-score variation of validation-datasets for various post-processing, including moderate, advanced, and GRU. To observe the performance variation across different network-complexities, the cross-database validation was performed with CNN models of (a) 2-layer, (b) 4-layer, (c) 8-layer, (d) 16-layer, (e) 32-layer, and (f) 64-layer depth. CNN model generated prediction-stream goes through post-processing, followed by the R-peak localisation where detection performance is calculated. 
  The models were trained using MIT-BIH-Arrhythmia and validated with the rest datasets.}
  \label{fig:gru_validations}
\end{figure*}

\begin{figure*}[h]
  \centering
  \begin{subfigure}[b]{0.32\textwidth}
    \includegraphics[width=\textwidth]{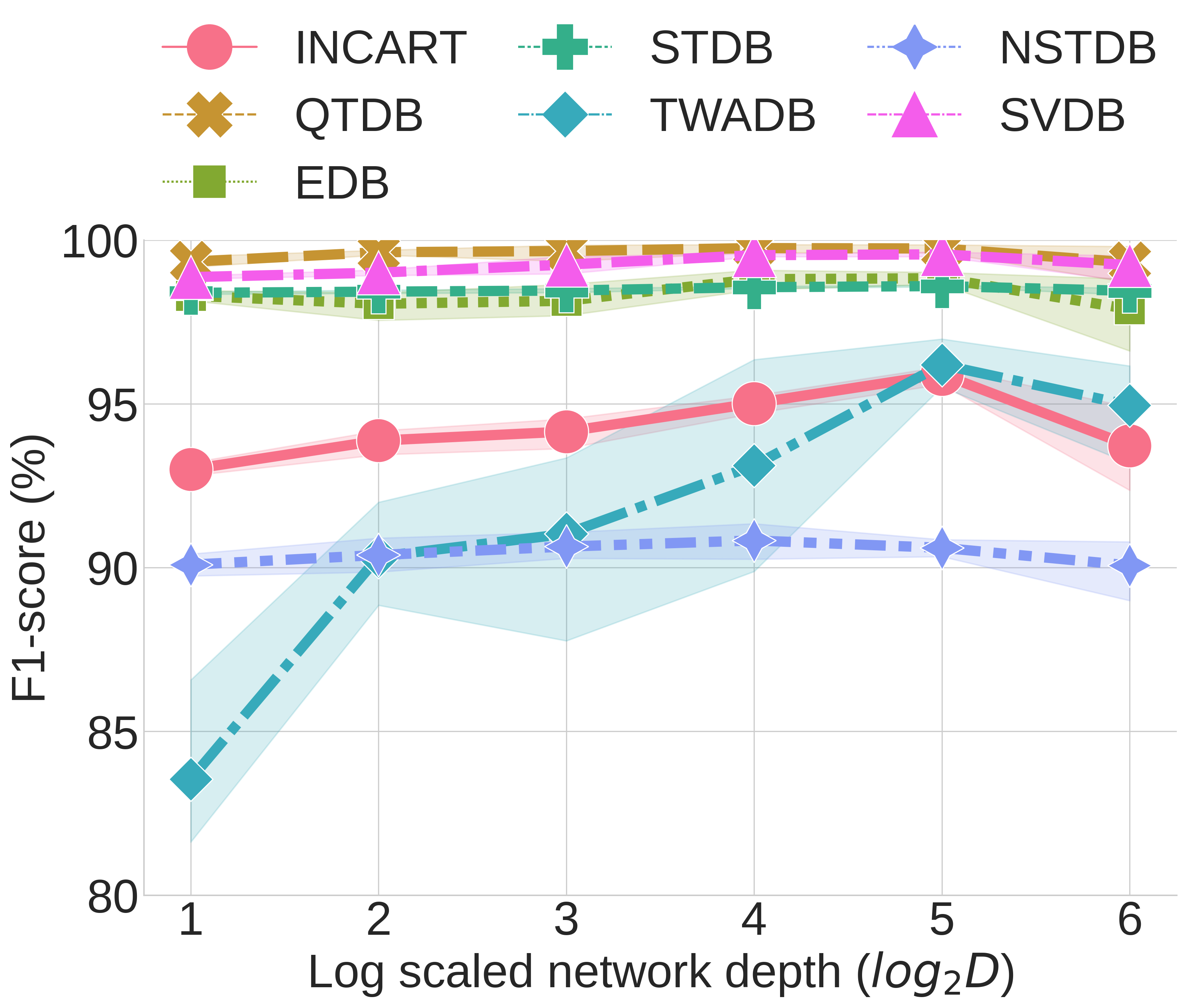}
    \caption{}
  \end{subfigure}
  \begin{subfigure}[b]{0.32\textwidth}
    \includegraphics[width=\textwidth]{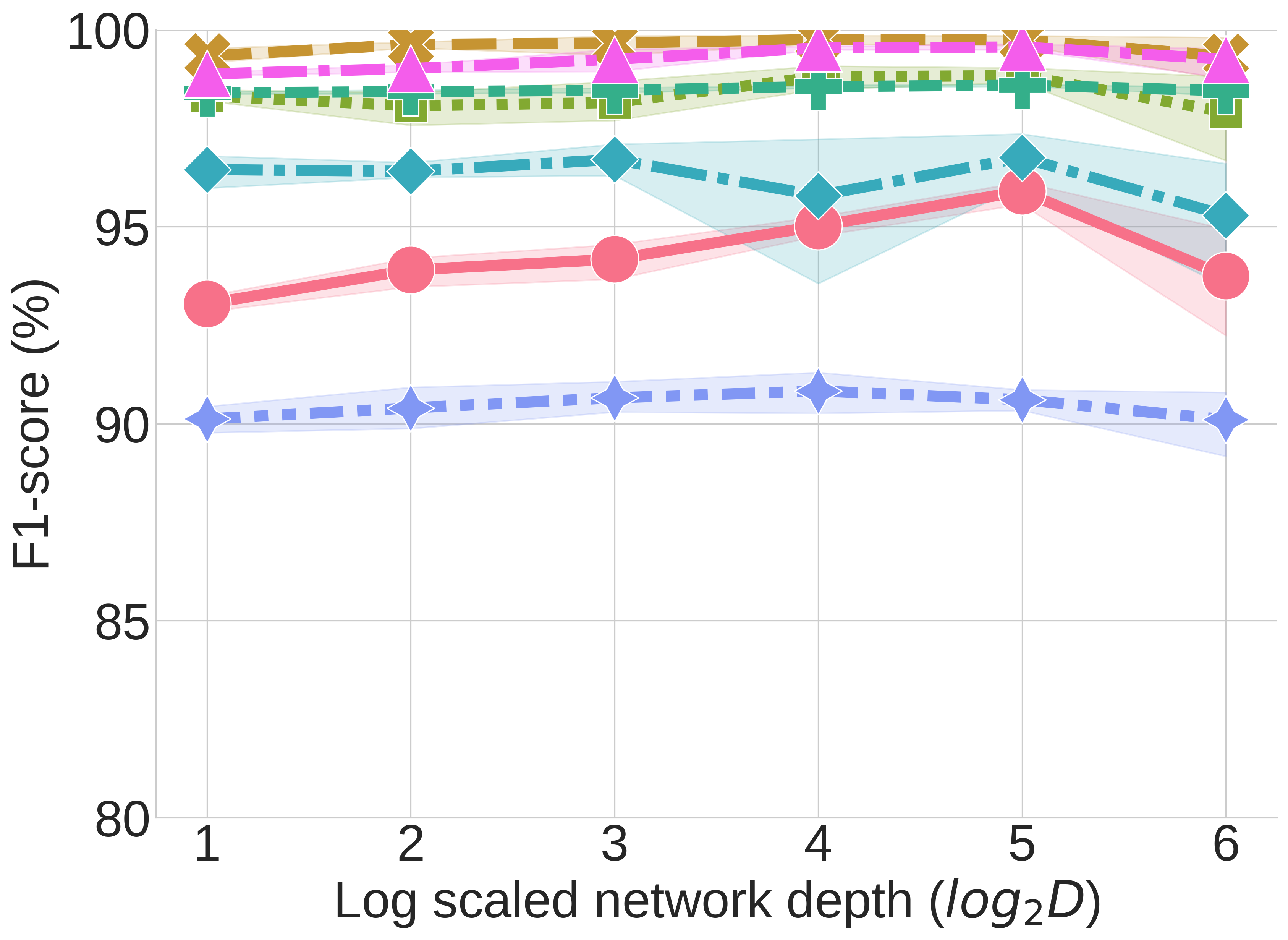}
    \caption{}
  \end{subfigure}
  \begin{subfigure}[b]{0.32\textwidth}
    \includegraphics[width=\textwidth]{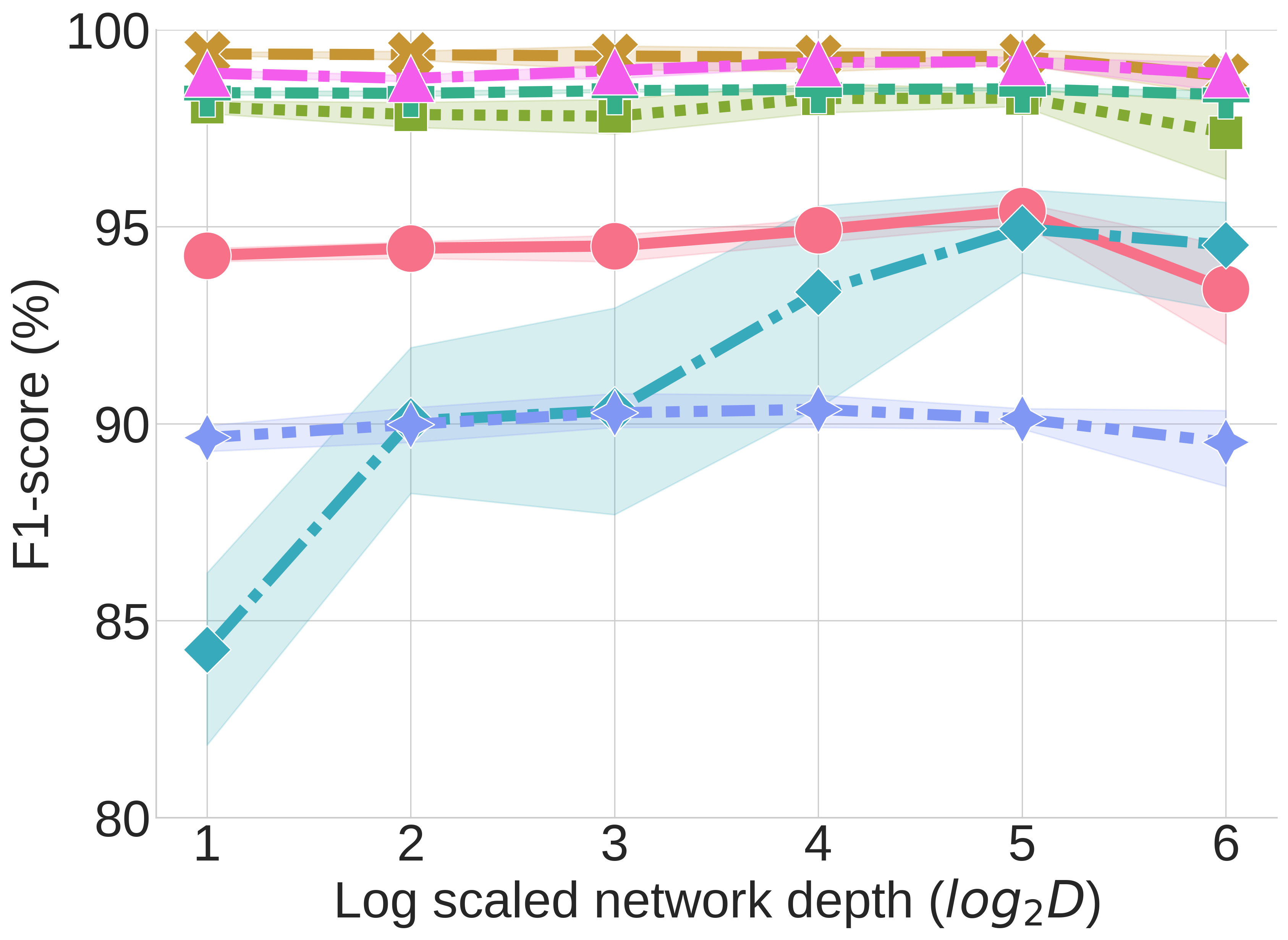}
    \caption{}
  \end{subfigure}

  \caption{F1-score variation for network-depths involving post-processing levels (a) moderate, (b) advanced, and (c) GRU-based. Network-depths of 2, 4, 8, 16, 32, and 64-layer are smoothed using $log_{2}D$, where D is the network-depth. The models were trained using MIT-BIH-Arrhythmia and validated with the rest datasets.}
  \label{fig:gru_validation_depths}
\end{figure*}

This study trains a GRU module to learn the post-processing so that the baseline-convnet and its deeper variants (i.e. CNN model with depth of 2, 4, 8, 16, 32, and 64-layer) generated prediction-streams for the records of validation-datasets can be refined to be used for R-peak localisation.
The GRU-based post-processing performance was compared with moderate and advanced post-processing.
The GRU-based approach was cross-validated with a broad range of validation-datasets, including, INCART, QT, EDB, STDB, TWADB, NSTDB, and SVDB.
The baseline-convnet and variants were trained with the MIT-BIH-Arrhythmia database, while the post-processing GRU was trained with only the baseline-convnet generated prediction-streams of the INCART subjects.
The MIT-BIH-Arrhythmia trained five CNN model instances per network-depth (produced by the CNN model training with five-fold internal cross-validation) were used to generate prediction-streams for each validation-dataset's subjects.

Figure~\ref{fig:gru_hyperparam}-a shows the GRU hyper-parameter space for the number of hidden-layers and sequence-length, encoded together as a double digit-number, where the first digit represents the hidden-layer number and the second accounts for sequence-length in seconds.
The figure shows possibilities of GRU hidden-layer and sequence-length combinations, including 11, 21, and 24, but selection of less-complex parameters (i.e. 11, GRU with a single hidden-layer and 1 second sequence-length) can be justified following Occam's Razor principle.
Regarding the choice of noisy prediction-streams for the GRU training, the Figure \ref{fig:gru_hyperparam}-b suggests that the training has insignificant impact for the noise-level of the input, and selection of the baseline-convnet (i.e. 2-layer CNN model) generated INCART prediction-streams are safe to go for GRU training.
The following test results were generated based on the optimised GRU which consists of a single hidden-layer and 1-second sequence-length.

Figure \ref{fig:gru_validations} shows comparative performance of post-processing levels, including the moderate, advanced, and GRU-based approaches across the validation-datasets of the baseline-convnet (i.e. 2-layer CNN model) and its deeper variants.
The baseline-convnet generated prediction-streams, followed by post-processing levels, as shown in Figure \ref{fig:gru_validations}-a, reveals the competitive scores of the moderate, advanced, and GRU post-processing, except for the TWADB, and INCART where the advanced and GRU post-processing were superior with around 12\% and 1.5\% margin respectively.

Figure \ref{fig:gru_validations}-b shows the 4-layer CNN model generated prediction-streams' performance and identifies a rapid development of moderate and GRU post-processing scores with more than 6\% margin for the TWADB.
As the network-depth increases, so does each post-processing level's F1-scores until the 32-layer depth (Figure \ref{fig:gru_validations}-e), beyond which all scores start to decrease (i.e. 64-layer network in Figure \ref{fig:gru_validations}-f).
With the single exception of TWADB, which almost equalizes in the 32-layer network and beyond. 
The GRU post-processing F1-scores have been closely following the moderate and advanced post-processing scores across the validation-datasets for all the network-depths.

Figure \ref{fig:gru_validation_depths} shows post-processing levels' performance variation with CNN model depths, which generate perdiction-streams to be forwarded to the post-processing step.
With moderate post-processing, in Figure \ref{fig:gru_validation_depths}-a, the F1-scores for all the datasets improved in different proportions, specifically, starting from the 4-layer network, the scores of all the datasets were 90\% or above, except for the NSTDB which were slightly below 90\% in 64-layer network.
The advanced post-processing, in Figure \ref{fig:gru_validation_depths}-b, was able to push all the dataset F1-scores at 90\% level or beyond from the baseline-convnet, which is a 2-layer CNN model.
The GRU post-processing, in Figure \ref{fig:gru_validation_depths}-c, follows a similar pattern of the moderate post-processing, but with TWADB showing slight improvements in the corresponding 2, 16, and 64-layer networks, and INCART being almost consistent from 2 to 32-layer networks.

Cross-database validation time-complexity depends on ECG signal duration of its subjects and the CNN model depth.
For example, INCART contains 75 subjects each of 30 minutes (i.e. 1800 seconds) length and considering 1-second overlapping with 3-seconds segments, on average, slightly lower than 1800 3-second segments went through the CNN model to generate prediction-streams.
For each INCART record, 2, 8, and 64-layer CNNs took on average 12 seconds of server time, however, Laptop CPU took on average 3, 9, and 50 seconds.
The server is equipped with Nvidia Tesla Volta V100-SXM2-32GB and the Laptop runs 1.90GHz Intel Core i7 8-core CPU with 16GB memory running Ubuntu 20.04.

\section{Discussion} \label{discussion}

This study categorised the post-processing, based on the QRS-detection literature, into moderate, and advanced levels (Algorithm \ref{algo:postprocessing_moderate}, and \ref{algo:postprocessing_advanced}).
A primitive salt-and-pepper filtration step (Algorithm \ref{algo:postprocessing_minimal}) was often required for these post-processing.
These levels include signal-processing tasks employing domain-specific knowledge to refine the CNN model generated binary prediction-stream, which was then forwarded for R-peak localisation.
There are certain threshold values, derived from domain-specific knowledge, including the minimum QRS-complex duration, and R-R interval, which were used in these post-processing levels to filter candidate QRS-complex.
ECG dataset diversity, especially, in terms of inter-patient variance and QRS-like false artifacts, may impose challenges to QRS-detectors to generalise, so the usage of threshold-based post-processing should be reduced as much as feasible.
A GRU, which is one kind of recurrent-neural-network, was trained to learn the post-processing by looking into the baseline-convnet generated prediction-streams of INCART subjects (four sets of prediction-streams) and compare with the corresponding annotation-streams.
The trained GRU was then used to post-process the baseline-convnet and it's deeper-variants' generated prediction-streams for the validation-datasets.

GRU post-processing was found to achieve similar performance of moderate and advanced post-processing, with the exception of TWADB, for which the advanced post-processing was superior, however, that margin decreased with the increase of CNN model-depth.
The GRU contribution is observable in the baseline-convnet for all the validation-datasets, as well as, in 4, 8, and 16-layer CNN model performance-scores (Figure \ref{fig:gru_validations}) with a gradually decreasing rate.
A particularly important observation regarding GRU post-processing is its ability to learn domain-specific parameter thresholds, including the minimum QRS-complex extent and R-R interval which were set to 64 milli-seconds and 200 milli-seconds equivalent number of samples respectively.

The TWADB was the single exception, which favored the advanced post-processing and achieved more than 95\% F1-score starting from shallow 2-layer baseline-convnet and this particularly reflects the strength of application of domain-specific knowledge for a QRS-detector in combining a shallow 2-layer CNN model with advanced post-processing.
On the other hand, the strength of GRU-based post-processing, probably is its ability to self-learn the temporal relationship of QRS-complex regions from the CNN model generated prediction-streams.
Although, the GRU post-processing failed to surpass the advanced post-processing, it marginally followed the advanced post-processing F1-scores (except the TWADB for 2 to 16-layer deep CNN models).
Moreover, the GRU for post-processing may be subject to further tuning, in terms of architecture (i.e. use of bi-directional or stacked GRUs may have possibilities) or training samples (i.e. use TWADB-like challenging datasets prediction-streams for training, instead of INCART), which could be an interesting future research.

CNN models with increased network-depths were found to be capable of generating strong enough prediction-streams for which the moderate-only post-processing was found to be adequate (Figure \ref{fig:gru_validations}-d,e,f).
This is particularly interesting to observe, in the QRS-detection literature using deep-learning-based approaches, where often a complex model is proposed to generate predictions along with post-processing which involves signal-processing with domain-specific knowledge \cite{cai_qrs_2020, liu_octave_2020}.
In our study, the strength of the CNN model's prediction-stream generation and post-processing were separately identified as modules, which, if carefully chosen, may help build a suitable solution deployable to either resource-constraint wearable devices or clinical settings.
A closer look in the Figure \ref{fig:gru_validation_depths} may reveal available solution-composing choices, for example, for over 90\% performance requirement, the options, including (i) 4-layer CNN model with moderate post-processing (Figure \ref{fig:gru_validation_depths}-a), (ii) 2-layer CNN model with advanced post-processing (Figure \ref{fig:gru_validation_depths}-b), or (iii) 4-layer CNN model with GRU post-processing with the appeal of being ignorant of domain-specific knowledge for post-processing, and rather depend on the GRU to learn these inherent parameters from the model prediction-streams itself.
Resource constraint devices may prefer the GRU-option since it uses a shallow 4-layer CNN (on average, it should not take more than 6 seconds to process 30 minutes-long INCART records with the CPU configuration, mentioned in the results section).

\begin{figure}[h]
  \centering
  \begin{subfigure}[b]{0.45\textwidth}
    \includegraphics[width=\textwidth]{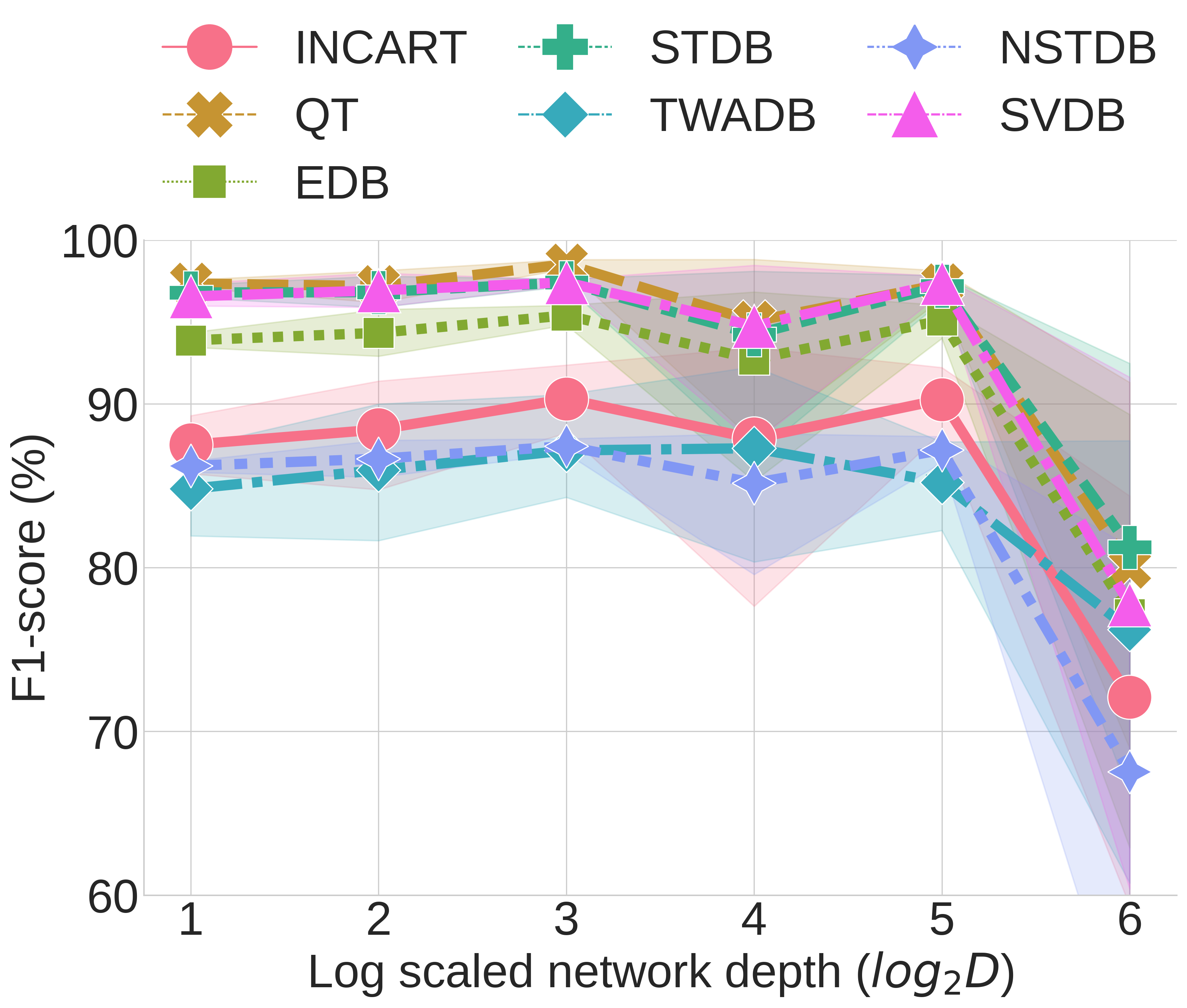}
  \end{subfigure}

  \caption{Cross-database F1-score response for network-depth changes of a CNN-GRU network where a single hidden-layer GRU module of 1-second sequence-length was placed just before the scoring-layer in Figure \ref{fig:network}. Each line represents each validation-dataset's response to network depths with the essential salt-and-pepper filtering. The model was trained using MIT-BIH-Arrhythmia and validated with the rest datasets.}
  \label{fig:cnn_gru_depths}
\end{figure}

The QRS-detection literature often advocates the usage of LSTM modules on top of feature-extracting CNN sub-networks to sequence convoluted high-level features \cite{yuen_inter-patient_2019}, however, in our study, the usage of a GRU on top of the baseline-convnet and its deeper variants did not yield better performance scores compared to the no-GRU counterparts, as shown in Figure~\ref{fig:cnn_gru_depths}.
The CNN-GRU shows comparatively stable performance for comparatively shallow CNN sub-networks (i.e. with 2, 4, and 8-layer CNN sub-network), but drops drastically for 64-layer CNN sub-network (with essential salt-and-pepper noise removal before the final R-peak localisation).
A particular configuration of CNN-GRU, yielded by optimisation for QRS-detection, may perform better. 
However, the observed general behavior of CNN-GRU (Figure~\ref{fig:cnn_gru_depths}) reveals that the GRU is sensitive to features obtained from different depths of CNN, while the GRU is stable with shallow CNN-features, which express more general information, but becomes more sensitive to deep CNN-features which express abstract-level information~\cite{chen2017ensemble}.
The GRU layer, intended to be used directly with the output binary prediction-stream, i.e., to refine QRS prediction-stream, can be argued to be exposed to comparatively less risk of being sensitive than its use just before the decision layer while dealing with abstract-level features.
Such an observation is an interesting deep-learning model design phenomena, however, it was least reflected in the literature, which requires further investigation.

\begin{figure}[h]
  \centering
  \begin{subfigure}[b]{0.41\textwidth}
    \includegraphics[width=\textwidth]{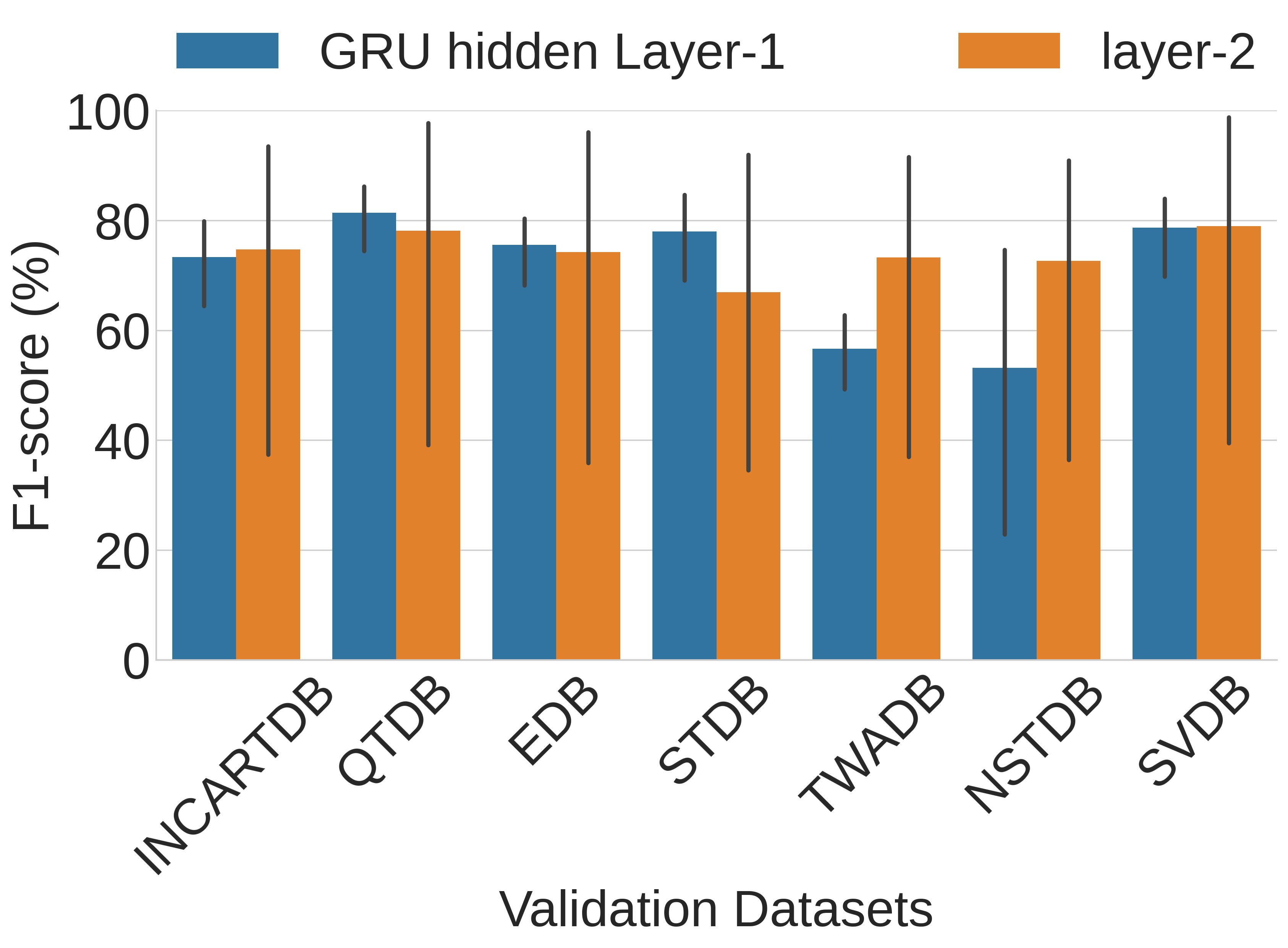}
  \end{subfigure}

  \caption{Cross-database F1-score response for validation-datasets using a GRU-only model with the essential salt-and-pepper filtering. A single and double hidden-layer GRUs were trained using MIT-BIH-Arrhythmia and validated with the rest datasets.}
  \label{fig:gru_depths}
\end{figure}

The use of an LSTM-only network is also available in QRS-detection literature \cite{laitala2020robust} to process an ECG-signal's morphology and temporal-sequences, however, a GRU-only network yielded poorer performance, in our experiment (Figure \ref{fig:gru_depths}), compared to the other two scenarios (CNN with manual post-processing; and CNN with GRU before (Figure~\ref{fig:cnn_gru_depths}) or after (Figure~\ref{fig:gru_validations},~\ref{fig:gru_validation_depths}) the CNN's decision layer).
It should be noted that the essential salt-and-pepper noise removal step was also used here, just before the final R-peak localisation.
We tried a single and double hidden-layer GRUs and the single hidden-layer GRU shows superiority to most of the validation-datasets, however, the double hidden-layer GRU was better for more challenging (i.e. TWADB) and noisy (i.e. NSTDB) datasets.
It can be argued that double-layer RNNs (as well as, stacked two single-layer RNNs) are more effective in noise-removal and understanding temporal relationship than the single-counterparts~\cite{han2021hybrid}.

CNN-only models performed better, starting from shallow 2 or 4-layer models, if post-processing with domain-specific heuristics were used, classified as moderate (Algorithm \ref{algo:postprocessing_moderate}), and advanced (Algorithm \ref{algo:postprocessing_advanced}) in this study.
The possibility of using a GRU module to learn post-processing from CNN model generated prediction-streams, being ignorant of any domain-specific parameters, was explored and found that GRUs are able to refine CNN model generated prediction-streams and yield performance that closely follows the moderate and advanced post-processing.
This may have potential, although GRU-based post-processing could not surpass the advanced post-processing, and subject to further investigation in order to be able to leverage the strength of CNN-model and post-processing (manual domain-knowledge-based or domain-agnostic GRU-based).

\section{Conclusion} \label{conclusion}

Post-processing has been an essential step in deep-learning-based QRS-detection studies in the literature, which often involves basic signal-processing tasks, including removal of isolated high or low predictions in the prediction-stream of an ECG record, as well as, tasks that involve domain-specific knowledge, including the minimum QRS-complex extent or R-R distance, to filter legitimate candidate QRS-complex for R-peak localisation.
The set of parameters, used to post-process QRS-complex prediction-streams, were found to assume different thresholds based on domain-specific knowledge that may favor study-specific datasets and may not generalise well across diverse range of datasets.
This study attempts to make a GRU (which is a variant of a recurrent-neural-network) learn the post-processing from CNN model-generated prediction-streams in order to avoid using domain-specific parameter-based post-processing for better generalisation.
The results suggest that GRU post-processing marginally follows the performance levels of moderate, and advanced post-processing.
In addition, by identifying individual strengths of a CNN model and post-processing, a modular QRS-detection solution design was suggested to combine each of them considering relative complexities for a target application deployment environment.


\section*{Acknowledgement}
This research was undertaken with the assistance of resources and services from the National Computational Infrastructure (NCI), which is supported by the Australian Government.

\bibliographystyle{IEEEtran}
\bibliography{main}

\begin{thebibliography}{10}
\providecommand{\url}[1]{#1}
\csname url@samestyle\endcsname
\providecommand{\newblock}{\relax}
\providecommand{\bibinfo}[2]{#2}
\providecommand{\BIBentrySTDinterwordspacing}{\spaceskip=0pt\relax}
\providecommand{\BIBentryALTinterwordstretchfactor}{4}
\providecommand{\BIBentryALTinterwordspacing}{\spaceskip=\fontdimen2\font plus
\BIBentryALTinterwordstretchfactor\fontdimen3\font minus
  \fontdimen4\font\relax}
\providecommand{\BIBforeignlanguage}[2]{{%
\expandafter\ifx\csname l@#1\endcsname\relax
\typeout{** WARNING: IEEEtran.bst: No hyphenation pattern has been}%
\typeout{** loaded for the language `#1'. Using the pattern for}%
\typeout{** the default language instead.}%
\else
\language=\csname l@#1\endcsname
\fi
#2}}
\providecommand{\BIBdecl}{\relax}
\BIBdecl

\bibitem{Kohler2002}
B.~U. Köhler, C.~Hennig, and R.~Orglmeister, ``The principles of software
  {{QRS}} detection.''

\bibitem{mehta2008svm}
S.~Mehta and N.~Lingayat, ``Svm-based algorithm for recognition of qrs
  complexes in electrocardiogram,'' \emph{Irbm}, vol.~29, no.~5, pp. 310--317,
  2008.

\bibitem{kropf2017ecg}
M.~Kropf, D.~Hayn, and G.~Schreier, ``Ecg classification based on time and
  frequency domain features using random forests,'' in \emph{2017 Computing in
  Cardiology (CinC)}.\hskip 1em plus 0.5em minus 0.4em\relax IEEE, 2017, pp.
  1--4.

\bibitem{saini2013qrs}
I.~Saini, D.~Singh, and A.~Khosla, ``Qrs detection using k-nearest neighbor
  algorithm (knn) and evaluation on standard ecg databases,'' \emph{Journal of
  advanced research}, vol.~4, no.~4, pp. 331--344, 2013.

\bibitem{habib_impact_2019}
\BIBentryALTinterwordspacing
A.~Habib, C.~Karmakar, and J.~Yearwood, ``Impact of {{ECG Dataset Diversity}}
  on {{Generalization}} of {{CNN Model}} for {{Detecting QRS Complex}},''
  vol.~7, pp. 93\,275--93\,285. [Online]. Available:
  \url{https://ieeexplore.ieee.org/document/8758818/}
\BIBentrySTDinterwordspacing

\bibitem{chandra_robust_2019}
\BIBentryALTinterwordspacing
B.~S. Chandra, C.~S. Sastry, and S.~Jana, ``Robust {{Heartbeat Detection From
  Multimodal Data}} via {{CNN}}-{{Based Generalizable Information Fusion}},''
  vol.~66, no.~3, pp. 710--717. [Online]. Available:
  \url{https://ieeexplore.ieee.org/document/8410035/}
\BIBentrySTDinterwordspacing

\bibitem{Xiang2018}
Y.~Xiang, Z.~Lin, and J.~Meng, ``Automatic {{QRS}} complex detection using
  two-level convolutional neural network,'' vol.~17, no.~1.

\bibitem{cai_qrs_2020}
\BIBentryALTinterwordspacing
W.~Cai and D.~Hu, ``{{QRS Complex Detection Using Novel Deep Learning Neural
  Networks}},'' vol.~8, pp. 97\,082--97\,089. [Online]. Available:
  \url{https://ieeexplore.ieee.org/document/9099511/}
\BIBentrySTDinterwordspacing

\bibitem{liu_octave_2020}
\BIBentryALTinterwordspacing
W.~Liu, X.~Wang, H.~Gao, C.~Yang, J.~Li, and C.~Liu, ``An {{Octave Convolution
  Neural Network}}-based {{QRS Detector}},'' in \emph{2020 {{International
  Conference}} on {{Sensing}}, {{Measurement}} \& {{Data Analytics}} in the Era
  of {{Artificial Intelligence}} ({{ICSMD}})}.\hskip 1em plus 0.5em minus
  0.4em\relax {IEEE}, pp. 413--418. [Online]. Available:
  \url{https://ieeexplore.ieee.org/document/9261658/}
\BIBentrySTDinterwordspacing

\bibitem{jia_high_2019}
\BIBentryALTinterwordspacing
M.~Jia, F.~Li, Z.~Chen, X.~Xiang, and X.~Yan, ``High {{Noise Tolerant
  R}}-{{Peak Detection Method Based}} on {{Deep Convolution Neural Network}},''
  vol. E102.D, no.~11, pp. 2272--2275. [Online]. Available:
  \url{https://www.jstage.jst.go.jp/article/transinf/E102.D/11/E102.D_2019EDL8097/_article}
\BIBentrySTDinterwordspacing

\bibitem{sarlija_convolutional_2017}
\BIBentryALTinterwordspacing
M.~Sarlija, F.~Jurisic, and S.~Popovic, ``A convolutional neural network based
  approach to {{QRS}} detection,'' in \emph{Proceedings of the 10th
  {{International Symposium}} on {{Image}} and {{Signal Processing}} and
  {{Analysis}}}.\hskip 1em plus 0.5em minus 0.4em\relax {IEEE}, pp. 121--125.
  [Online]. Available: \url{http://ieeexplore.ieee.org/document/8073581/}
\BIBentrySTDinterwordspacing

\bibitem{Lee2018}
J.~S. Lee, M.~Seo, S.~W. Kim, and M.~Choi, ``Fetal {{QRS}} detection based on
  convolutional neural networks in noninvasive fetal electrocardiogram,''
  vol.~4, pp. 75--78.

\bibitem{LEE201966}
\BIBentryALTinterwordspacing
J.~S. Lee, S.~J. Lee, M.~Choi, M.~Seo, and S.~W. Kim, ``{{QRS}} detection
  method based on fully convolutional networks for capacitive
  electrocardiogram,'' vol. 134, pp. 66--78. [Online]. Available:
  \url{https://www.sciencedirect.com/science/article/pii/S0957417419303628}
\BIBentrySTDinterwordspacing

\bibitem{goldberger_physiobank_2000}
\BIBentryALTinterwordspacing
A.~L. Goldberger, L.~A.~N. Amaral, L.~Glass, J.~M. Hausdorff, P.~C. Ivanov,
  R.~G. Mark, J.~E. Mietus, G.~B. Moody, C.-K. Peng, and H.~E. Stanley,
  ``{{PhysioBank}}, {{PhysioToolkit}}, and {{PhysioNet}}: {{Components}} of a
  {{New Research Resource}} for {{Complex Physiologic Signals}},'' vol. 101,
  no.~23. [Online]. Available:
  \url{https://www.ahajournals.org/doi/10.1161/01.CIR.101.23.e215}
\BIBentrySTDinterwordspacing

\bibitem{moody_impact_2001}
\BIBentryALTinterwordspacing
G.~Moody and R.~Mark, ``The impact of the {{MIT}}-{{BIH Arrhythmia
  Database}},'' vol.~20, no.~3, pp. 45--50, May-June/2001. [Online]. Available:
  \url{http://ieeexplore.ieee.org/document/932724/}
\BIBentrySTDinterwordspacing

\bibitem{laguna_database_1997}
\BIBentryALTinterwordspacing
P.~Laguna, R.~Mark, A.~Goldberg, and G.~Moody, ``A database for evaluation of
  algorithms for measurement of {{QT}} and other waveform intervals in the
  {{ECG}},'' in \emph{Computers in {{Cardiology}} 1997}.\hskip 1em plus 0.5em
  minus 0.4em\relax {IEEE}, pp. 673--676. [Online]. Available:
  \url{http://ieeexplore.ieee.org/document/648140/}
\BIBentrySTDinterwordspacing

\bibitem{taddei_european_1992}
\BIBentryALTinterwordspacing
A.~Taddei, G.~Distante, M.~Emdin, P.~Pisani, G.~B. Moody, C.~Zeelenberg, and
  C.~Marchesi, ``The {{European ST}}-{{T}} database: Standard for evaluating
  systems for the analysis of {{ST}}-{{T}} changes in ambulatory
  electrocardiography,'' vol.~13, no.~9, pp. 1164--1172. [Online]. Available:
  \url{https://academic.oup.com/eurheartj/article/451781/The}
\BIBentrySTDinterwordspacing

\bibitem{moody_noise_1984}
G.~B. Moody, W.~Muldrow, and R.~G. Mark, ``A noise stress test for arrhythmia
  detectors,'' vol.~11, no.~3, pp. 381--384.

\bibitem{greenwald_improved_1990}
S.~Greenwald, ``Improved detection and classification of arrhythmias in
  noise-corrupted electrocardiograms using contextual information.''

\bibitem{ajdaraga_analysis_2017}
\BIBentryALTinterwordspacing
E.~Ajdaraga and M.~Gusev, ``Analysis of sampling frequency and resolution in
  {{ECG}} signals,'' in \emph{2017 25th {{Telecommunication Forum}}
  ({{TELFOR}})}.\hskip 1em plus 0.5em minus 0.4em\relax {IEEE}, pp. 1--4.
  [Online]. Available: \url{http://ieeexplore.ieee.org/document/8249438/}
\BIBentrySTDinterwordspacing

\bibitem{proakis_digital_2004}
J.~G. Proakis and D.~G. Manolakis, \emph{Digital Signal Processing}.\hskip 1em
  plus 0.5em minus 0.4em\relax {PHI publication}.

\bibitem{guo_inter-patient_2019}
\BIBentryALTinterwordspacing
L.~Guo, G.~Sim, and B.~Matuszewski, ``Inter-patient {{ECG}} classification with
  convolutional and recurrent neural networks,'' vol.~39, no.~3, pp. 868--879.
  [Online]. Available:
  \url{https://linkinghub.elsevier.com/retrieve/pii/S0208521618304200}
\BIBentrySTDinterwordspacing

\bibitem{navab_u-net_2015}
\BIBentryALTinterwordspacing
O.~Ronneberger, P.~Fischer, and T.~Brox, ``U-{{Net}}: {{Convolutional
  Networks}} for {{Biomedical Image Segmentation}},'' in \emph{Medical {{Image
  Computing}} and {{Computer}}-{{Assisted Intervention}} – {{MICCAI}} 2015},
  ser. Lecture {{Notes}} in {{Computer Science}}, N.~Navab, J.~Hornegger, W.~M.
  Wells, and A.~F. Frangi, Eds.\hskip 1em plus 0.5em minus 0.4em\relax
  {Springer International Publishing}, vol. 9351, pp. 234--241. [Online].
  Available: \url{http://link.springer.com/10.1007/978-3-319-24574-4_28}
\BIBentrySTDinterwordspacing

\bibitem{pelt_mixed-scale_2018}
\BIBentryALTinterwordspacing
D.~M. Pelt and J.~A. Sethian, ``A mixed-scale dense convolutional neural
  network for image analysis,'' vol. 115, no.~2, pp. 254--259. [Online].
  Available: \url{http://www.pnas.org/lookup/doi/10.1073/pnas.1715832114}
\BIBentrySTDinterwordspacing

\bibitem{ioffe2015batch}
S.~Ioffe and C.~Szegedy, ``Batch normalization: Accelerating deep network
  training by reducing internal covariate shift,'' in \emph{International
  conference on machine learning}.\hskip 1em plus 0.5em minus 0.4em\relax PMLR,
  2015, pp. 448--456.

\bibitem{cho2014learning}
K.~Cho, B.~Van~Merri{\"e}nboer, C.~Gulcehre, D.~Bahdanau, F.~Bougares,
  H.~Schwenk, and Y.~Bengio, ``Learning phrase representations using rnn
  encoder-decoder for statistical machine translation,'' \emph{arXiv preprint
  arXiv:1406.1078}, 2014.

\bibitem{yuen_inter-patient_2019}
\BIBentryALTinterwordspacing
B.~Yuen, X.~Dong, and T.~Lu, ``Inter-{{Patient CNN}}-{{LSTM}} for {{QRS Complex
  Detection}} in {{Noisy ECG Signals}},'' vol.~7, pp. 169\,359--169\,370.
  [Online]. Available: \url{https://ieeexplore.ieee.org/document/8911411/}
\BIBentrySTDinterwordspacing

\bibitem{chen2017ensemble}
J.~Chen, Y.~Wang, Y.~Wu, and C.~Cai, ``An ensemble of convolutional neural
  networks for image classification based on lstm,'' in \emph{2017
  International Conference on Green Informatics (ICGI)}.\hskip 1em plus 0.5em
  minus 0.4em\relax IEEE, 2017, pp. 217--222.

\bibitem{laitala2020robust}
J.~Laitala, M.~Jiang, E.~Syrj{\"a}l{\"a}, E.~K. Naeini, A.~Airola, A.~M.
  Rahmani, N.~D. Dutt, and P.~Liljeberg, ``Robust ecg r-peak detection using
  lstm,'' in \emph{Proceedings of the 35th annual ACM symposium on applied
  computing}, 2020, pp. 1104--1111.

\bibitem{han2021hybrid}
S.~Han, Z.~Meng, X.~Zhang, and Y.~Yan, ``Hybrid deep recurrent neural networks
  for noise reduction of mems-imu with static and dynamic conditions,''
  \emph{Micromachines}, vol.~12, no.~2, p. 214, 2021.

\end{thebibliography}


\end{document}